# Unknown Biases and Timing Constraints in Timed Automata

Darion Haase  and Joost-Pieter Katoen

Software Modeling and Verification Group
RWTH Aachen University, Aachen, Germany[*]
{darion.haase,katoen}@cs.rwth-aachen.de

## Abstract

Timed automata are the formal model for real-time systems. Extensions with discrete probabilistic branching have been considered in the literature and successfully applied. Probabilistic timed automata (PTA) do require all branching probabilities and clock constraints to be constants. This report investigates PTA in which this constraint is relaxed: both branching probabilities and clock constraints can be parametric. We formally define this PTA variant and define its semantics by an uncountable parametric Markov Decision Process (pMDP). We show that reachability probabilities in parametric L/U-PTA can be reduced to considering PTA with only parametric branching probabilities. This enables the usage of existing techniques from the literature. Finally, we generalize the symbolic backward and digital clock semantics of PTA to the setting with parametric probabilities and constraints.

## 1 Introduction

**Timed automata.** Timed automata (TA) [2, 24] are the formal model that is used for the description and automated analysis of real-time systems. Such automata use a set of real-valued clocks which all implicitly progress at the same speed. Constraints on the clock values (comparisons of clocks to natural numbers) are used to describe when transitions are enabled. For instance, a transition with constraint $c \leq 3$ for clock $c$ asserts that this transition is only possible if the clock $c$ is at most three. Such constraints can also be used to restrict the amount of time that can be spent in a location. Together with transition guards, these location invariants can enforce a transition to be taken at a given time point. Clocks cannot be manipulated; they can only be reset when taking a transition. Despite that the reachability problem—can a given location be reached within a deadline?—is **PSPACE**-complete, TA can efficiently be analysed due to unremitting developments in algorithms, symbolic data structures and aggressive abstraction techniques. The key is that the uncountable concrete state space (due to the real-valued clocks) of a TA can be abstracted into a finite abstract-time transition system. UPPAAL [15] is the most popular software tool for model checking of TA.

---

[*]This work is supported by the DFG Project 662562 PASIWY.

**Probabilistic timed automata.** An extension with *probabilistic branching* has been proposed in [62]. This enables to model uncertainty such as the loss probability of a wireless communication channel in the behaviour of real-time systems. Transitions in probabilistic TA (PTA) do not result in a single target location but rather in a discrete distribution over target locations. In addition, different branches may involve different clock resets. Guards (i.e., clock constraints) are however in common for all branches of a transition. The model-checking objective now shifts from a purely qualitative analysis—can a given location be reached within a given deadline? —to a quantitative analysis: is the probability to reach a given location within a deadline at least a half?

The concrete semantics of a PTA is an uncountable Markov decision process (MDP) [69]. As shown in [62], it suffices to consider a finite abstraction of this process, i.e., a finite MDP, to obtain extremal reachability probabilities. As described in [68], there are two main practical techniques to obtain such abstractions: a backwards procedure and a digital clock approach. The former starts from the reachability target and aims to compute a finite MDP in a symbolic manner by considering zones. The latter takes the perspective that the only relevant clock values are natural. This view is correct in the absence of strict clock constraints. PRISM [56] and MCSTA [39] are two software tools that can be used to compute such probabilities.

**The need for parameters.** While the use of probabilities enables modelling uncertainty in real-time systems, all probability values need to be concrete values such as $\frac{1}{2}$, $\frac{3}{4}$ etc. In several circumstances, such concrete values are not at our disposal. A similar issue occurs when defining the clock constraints: the natural numbers with which the clocks are compared (such as time-out values) need to be all known in advance. The specific choice of time-out threshold may influence the validity of a (safety) property. Rather than checking a reachability probability for a given concrete time-out value and a concrete loss probability of the communication medium, it is of interest to synthesize for which time-out values and loss probabilities a given reachability probability can at least be guaranteed.

**Parametric versions of (P)TA.** To address this, variants of TA with parametric clock constraints [3, 45], and, more recently, a parametric version of PTA has been proposed [40] with parametric branching probabilities. Parametric TA enable to write constraints in which clocks are compared against parameters, e.g. $c \leq T$ (where $T$ is an unknown). This approach has been extended to PTA in [47]. In parametric PTA [40], branching probabilities such as $p$ and $1-p$ can occur for unknown $0 < p < 1$. In fact, rational functions such as $\frac{p}{2} + \frac{1}{4}$ are allowed, as long as a valuation of $p$ results in a probability.

The analogue to the verification objectives in TA and PTA, their parametric variants, naturally lead to synthesis questions such as: what is the minimal time-out value for which a location (modelling that all packets are successfully transmitted) is reached with a high probability? Or, what is the maximal loss probability that a communication channel can admit while ensuring that all packets are transferred with a probability of, say, at least 90%? These synthesis questions can be considered as a kind of design-space exploration.

**Focus of this work.** This technical report considers the natural follow-up: *parametric clock constraints in combination with parametric probabilistic branching*. Thus, we consider unknown time-out values such as $T$ in the presence of parametric branching such as $p$ and $1-p$ modelling unknown failure rates. We consider both parameter 'types' as disjoint sets, e.g., we do not allow constraints such as $c < N$ together with branching probability $\frac{1}{N}$. Thus: some parameters only occur in clock constraints, whereas others only occur in probabilistic branching. Note that parameters can occur at various places, e.g., in multiple transition guards or in multiple random choices. This can lead to trade-offs: for certain transitions a small value of a parameter maximizes the reachability probability, whereas for others a large value is profitable. The same applies to the probability parameters. Our approach extends the work on parametric PTA [40] with parametric clock constraints.

**Our model and objectives.** This report defines *probability-parametric clock-parametric probabilistic timed automata*, called p$\mathbb{P}$pTA, which allow simultaneous parametrization of the transition probabilities and the clock constraints. We provide a formal semantics of this model in terms of parametric MDPs [52]. We use parametric kernels from measure theory to rigorously define this. States of such pMDPs are tuples consisting of a location, the clock values, and the parameter values. To the best of our knowledge, the TA model with these two kinds of parameters has not been formalized before. As objective, we consider finding the clock and/or probabilistic variable valuations that maximize the maximal reachability probability of a given location within a deadline. Whereas for a PTA, one considers the maximal reachability probability for a given set of constant values in the clock constraints and probabilistic choices, we consider an *optimization* problem over all possible values of parameters. Dually to the maximization problem, we consider the minimization over all parameters for minimal reachability probabilities.

**L/U timed automata with probabilistic parameters.** Timed automata with parametric clock constraints are hard. Several elementary objectives on these timed automata are undecidable [3, 4] such as the emptiness problem. L/U timed automata [45] are an elegant model that achieves a good compromise between expressive power and decidability. A given clock constraint parameter can either only occur as a lower bound on a clock, or as an upper bound. For example, it cannot occur as lower bound for some transition and as upper bound in some location invariant. In contrast to the general case, the emptiness problem for L/U timed automata is decidable. We study the extension of *L/U timed automata with probabilistic parameters.* We establish that—under mild restrictions—for this model the clock parameters are irrelevant to find optimal parameter values for maximal reachability probabilities. A direct consequence of this result is that existing techniques and tool-support for parametric PTA [40] can be readily used. Clock constraint parameters thus come for free!

**Abstractions.** We then explore two approaches to handle the uncountable state space of the p$\mathbb{P}$pTA model. Note that the uncountable nature no longer stems from the real-valued clocks, but also by the real-valued probabilistic parameters (the clock constraint parameters have a natural value). We consider two approaches that successfully handle the uncountable state space through abstraction for non-parametric models such as TA and PTA. Specifically, we generalize the *backwards exploration* approach [64] for PTA that collapses time-elapse transitions into symbolic states that (using the concept of zones) represent multiple valuations of clocks at once. The other approach we extend is the *digital clock* semantics [43] which is based on the idea that in a TA without strict inequalities, dense real-time steps are not needed, and integer time steps suffice. We show that both the backward reachability and digital clock approaches readily extend to the probability and clock parametric setting and can be used for reachability analysis in different ways. This result can be seen as a generalization of the results for parametric PTA in [40].

**Related work.** We follow a similar approach as [40] for probability-parametric probabilistic timed automata in extending different known semantics from the traditional non-parametric setting. Similar results were first presented in [13].

In [71] a version of probabilistic timed automata in which the probabilities can depend on the clocks' current values was introduced. The interdependencies between clocks and transition probabilities allow to model more complicated situations, but also make it harder to transfer approaches from existing models to this formalism. Inherent to our formalization of pPpTA is a strict separation of probabilistic and timed behaviour. Importantly, the parameters for probabilities and clock constraints are separate and put in no relation to each other. This allows for analysis of either parametric aspect independent of the other.

A related concept of parametric interval probabilistic timed automata [6] does allow each transition probability to be specified as an interval from which a concrete value can be picked. In particular, this does not allow to model interdependencies between probabilities of different transitions which can be achieved with our formalism by using the same parameter in both places. In addition, their work is focused on the general question of whether a proper instantiation of such a timed automaton is possible, while we assume to be (implicitly) provided with a set of valid parameter valuations.

**Outline.** Section 2 presents the necessary preliminaries, in particular on measure theory, parametric Markov kernels and parametric Markov Decision Processes (pMDPs). Section 3 defines probability-parametric clock-parametric probabilistic timed automata (pPpTA), and Section 4 presents their semantics as uncountable pMDPs and the optimal reachability parameter synthesis problems we consider. Section 5 considers the subclass of pPpTAs in the form of L/U-automata for which finding optimal parameter valuations is decidable. We show that an analogous extension to the quantitative case can be achieved. Section 6 considers alternative symbolic semantics: backwards semantics and digital clock semantics, for pPpTAs. We show that both approaches can be applied to the parametric setting, in the same spirit as [40].

## 2 Preliminaries

### 2.1 Notation

We write $\mathbb{N}$ for the set of natural numbers, $\mathbb{N}_0 = \mathbb{N} \cup \{0\}$ for the natural numbers with zero, $\mathbb{Z}$ for the set of all integers and $\mathbb{R}$ for the set of all real numbers. We use $\mathbb{R}_{\geq 0} := \{x \in \mathbb{R} \mid x \geq 0\}$ for the set of non-negative reals and similar notation for $\leq, >, \cdots$ and other sets, e.g. $\mathbb{N} = \mathbb{Z}_{>0}$. For $a, b \in \mathbb{Z}$ we write $[a, b] := \{x \in \mathbb{Z} \mid a \leq x \leq b\}$ and $[a, \infty) = [a, \infty] := \{x \in \mathbb{Z} \mid a \leq x\}$. For $a, b \in \mathbb{R}$ we write $[a, b]_\mathbb{R} := \{x \in \mathbb{R} \mid a \leq x \leq b\}$ for the closed interval of numbers between $a$ and $b$ inclusive. Similarly we will write $[a, b)_\mathbb{R}, (a, b]_\mathbb{R}, (a, b)_\mathbb{R}$ for the half open and open intervals. Note the subscript used for the real intervals. Intervals without subscript will always refer to the integer intervals over $\mathbb{Z}$.

For an (index) set $I$ and a set $A$, a family in $A$ over $I$ is a map $I \to A, i \mapsto a_i$ written as $(a_i)_{i \in I}$. The set of all families in $A$ over $I$ resp. maps from $I$ to $A$ is denoted by $A^I$. A sequence is a family over $\mathbb{N}_0$ or $\mathbb{N}$, a finite sequence (or tuple) is a family over $[1, n]$ or $[0, n]$ for some $n \in \mathbb{N}_0$.

The image of a map $f: A \to B$ is $\operatorname{Im}(f) := f(A) = \{ f(a) \mid a \in A \}$. For a map $f: A \to B$ and $A' \subseteq A$ we let the restriction of $f$ to $A'$ be $f\!\restriction_{A'}: A' \to B, a \mapsto f(a)$. For singletons $a \in A$ we will also write $f\!\restriction_a$ for $f\!\restriction_{\{a\}}$. This notation is lifted to sets $F \subseteq B^A$ element-wise: $F\!\restriction_{A'} := \{ f\!\restriction_{A'} \mid f \in F \}$. A partial map $f: A \rightharpoonup B$ is a map $f: A' \to B$ defined on some subset $\operatorname{dom}(f) := A' \subseteq A$ called the domain of $f$.

A set $J \subseteq I$ is called separable in $F \subseteq A^I$ if $F = F\!\restriction_J \times F\!\restriction_{I \setminus J}$, with the appropriate reordering of entries.

Two sets $A$ and $B$ are disjoint, if $A \cap B = \emptyset$. We will write $A \,\dot\cup\, B$ to indicate the union $A \cup B$ of two disjoint sets $A$ and $B$. For a set $X$, $\mathcal{P}(X) := \{ X' \mid X' \subseteq X \}$ denotes its power set.

For functions $f: A \to B$ and $f': A' \to B'$ with disjoint sets $A$ and $A'$, we define the element-wise union $f \,\dot\cup\, f': A \,\dot\cup\, A' \to B \cup B'$, with

$$x \mapsto \begin{cases} f(x), & \text{if } x \in A, \\ f'(x), & \text{if } x \in A'. \end{cases}$$

For $f: A \to B, a \in A, b \in B$ we let $f[a \mapsto b]$ denote the map which behaves like $f$ but maps $a$ to $b$, i.e. $f[a \mapsto b] = f\!\restriction_{A \setminus \{a\}} \,\dot\cup\, (a \mapsto b)$.

## 2.2 Measure Theory

To properly define the semantics of probabilistic timed automata with probability and clock parameters, we need to employ measure theory. We present the necessary measure theoretic concepts based on [22, 34]. Assuming the reader is familiar with basic concepts from probability theory, we mention the corresponding terminology from probability theory to make the definitions easier to understand.

**Definition 2.1** ($\sigma$-algebra)**.** Given a set $X$. A *$\sigma$-algebra* (of subsets) of $X$ is a collection $\Sigma$ of subsets of $X$ with

- $\Sigma$ contains the empty set: $\emptyset \in \Sigma$,

- $\Sigma$ is closed under complements: for all $E \in \Sigma$, also $X \setminus E \in \Sigma$,

- $\Sigma$ is closed under countable unions: if $(E_i)_{i \in \mathbb{N}}$ is a family in $\Sigma$, then $\bigcup_{i \in \mathbb{N}} E_i \in \Sigma$.

The elements $E \in \Sigma$ are called measurable sets. A pair $(X, \Sigma)$ with $\Sigma$ $\sigma$-algebra of $X$ is called a measurable space.

In the setting of probability theory, the measurable sets $E \in \Sigma$ correspond to the *events* which a *probability distribution* assigns a probability to. A standard way to obtain a $\sigma$-algebra is by extending a given set such that it fulfills the axioms of a $\sigma$-algebra.

**Definition 2.2.** Let $X$ be a set and $G$ be a collection of subsets of $X$. The *$\sigma$-algebra generated by $G$*, denoted $\sigma(G)$, is the smallest $\sigma$-algebra containing $G$.

On countable sets $X$, the standard $\sigma$-algebra is the power set $\mathcal{P}(X)$, which is precisely the $\sigma$-algebra generated by the collection of all singleton sets: $\mathcal{P}(X) = \sigma(\{\,\{\,x\,\} \mid x \in X\,\})$. On uncountable sets, most prominently the real numbers $\mathbb{R}$, the $\sigma$-algebra generated by the singletons is too small to be useful, for example it does not contain the open intervals $(x,y)_{\mathbb{R}}$. Instead, the most commonly used $\sigma$-algebra used on $\mathbb{R}^n$ is the Borel $\sigma$-algebra:

**Definition 2.3.** The *Borel $\sigma$-algebra* $\mathcal{B}(\mathbb{R}^n)$ is the $\sigma$-algebra generated by all open sets on $\mathbb{R}^n$. For $X \subseteq \mathbb{R}^n$ we define the $\sigma$-algebra $\mathcal{B}(X) \coloneqq \{\, X \cap E \mid E \in \mathcal{B}(\mathbb{R}^n)\,\}$.

With $\sigma$-algebras generalizing events from probability theory, we need a suitable generalization of probability distributions:

**Definition 2.4** (Measure)**.** Given a measurable space $(X, \Sigma_X)$. A (positive, finite) *measure* on $(X, \Sigma_X)$ is a function $\mu\colon \Sigma_X \to \mathbb{R}_{\geq 0}$, such that:

- $\mu(\emptyset) = 0$,
- For each family $(E_i)_{i \in \mathbb{N}}$ of pairwise disjoint sets $E_i \in \Sigma_X$: $\mu(\dot{\bigcup}_{i \in \mathbb{N}} E_i) = \sum_{i \in \mathbb{N}} \mu(E_i)$.

We let $|\mu| \coloneqq \mu(X)$ denote the size or norm of measure $\mu$. $\mu$ is called a subprobability measure, if $|\mu| \leq 1$, and a probability measure, if $|\mu| = 1$. We write $0$ for the measure that has the value $0$ for every measurable set.

We let $\mathbb{M}(\Sigma_X)$, or simply $\mathbb{M}(X)$ if $\Sigma_X$ is clear from the context, denote the set of all measures on $(X, \Sigma_X)$. A triple $(X, \Sigma_X, \mu)$ is called a measure space.

Measures are closed under (positive) linear combinations: $a_1\mu_1 + a_2\mu_2$ is a measure for $a_i \in \mathbb{R}_{\geq 0}$, and measures $\mu_i$. Since we only consider finite measures $\mu$, we can easily transform them (except $\mu = 0$) to probability measures by rescaling them: $\frac{1}{|\mu|}\mu$. We will thus also use the term *distributions* to refer to measures. The most basic example of a measure is the (probability) Dirac-measure of $x \in X$:

$$\mathfrak{d}_x\colon \Sigma_X \to \mathbb{R}_{\geq 0}, E \mapsto \begin{cases} 1, & \text{if } x \in E, \\ 0, & \text{if } x \notin E. \end{cases}$$

**Definition 2.5.** Given a measure space $(X, \Sigma_X, \mu)$. A property (of a function, relation, ...) on $X$ holds $\mu$-*almost surely* ($\mu$-*a.s.*) if there is a $\mu$-null set $N \in \Sigma_X, \mu(N) = 0$ such that the property holds on $X \setminus N$.

Measurable functions generalize the concept of *random variables* from probability theory.

**Definition 2.6** (Measurable function)**.** Given measurable spaces $(X, \Sigma_X)$ and $(Y, \Sigma_Y)$. A $(\Sigma_X, \Sigma_Y)$-*measurable function* is a function $f\colon X \to Y$ with $f^{-1}(E_Y) \in \Sigma_X$ for all $E_Y \in \Sigma_Y$. If $\Sigma_X$ and $\Sigma_Y$ are clear from the context we say $f$ is a measurable function.

The most commonly considered random variables which assign real values to the events correspond to $(\Sigma_X, \mathcal{B}(\mathbb{R}))$-measurable functions. For a measurable set $E \in \Sigma_X$, the indicator or characteristic function

$$\mathbb{1}_E\colon X \to \mathbb{R}_{\geq 0}, x \to \begin{cases} 1, & \text{if } x \in E, \\ 0, & \text{if } x \notin E, \end{cases}$$

is a measurable function. Note that $\mathfrak{d}_x(E) = \mathbb{1}_E(x)$.

**Proposition 2.7.** *Given measurable spaces $(X_i, \Sigma_{X_i}), i \in [1,3]$.*

- *The composition of measurable functions $f_1: X_1 \to X_2, f_2: X_2 \to X_3$ is a measurable function $f_2 \circ f_1: X_1 \to X_3$.*
- *Combinations on $\mathbb{R}$: if $f_1, f_2: X_1 \to \mathbb{R}$ are measurable functions, $a_1, a_2 \in \mathbb{R}$ then $a_1 f_1 + a_2 f_2$ and $f_1 \cdot f_2$ are also measurable functions.*

Measurable functions can be used to transform measures to a different measurable space:

**Definition 2.8** (Push-forward measure). Given measurable spaces $(X, \Sigma_X)$ and $(Y, \Sigma_Y)$ and a $(\Sigma_X, \Sigma_Y)$-measurable function $f: X \to Y$. For a measure $\mu: \Sigma_X \to \mathbb{R}_{\geq 0}$ the *push-forward measure of $\mu$ under $f$* is defined as

$$f\#\mu := \mu \circ f^{-1}: \Sigma_Y \to \mathbb{R}_{\geq 0}.$$

**Lebesgue Integration**

Lebesgue integration provides another way to combine a measurable function $f: X \to \mathbb{R}_{\geq 0}$ with a measure $\mu: \Sigma_X \to \mathbb{R}_{\geq 0}$. Intuitively, it corresponds to the $\mu$-weighted sum of $f$ over $X$: $\sum_{x \in X} f(x) \cdot \mu(x)$. From the point of probability theory this corresponds to the expected value of the random variable $f$ under the distribution $\mu$. We omit detailed definitions and refer to [22, 34] for a thorough explanation of Lebesgue integrals. We will denote the Lebesgue integral of $f$ over $\mu$ as

$$\int_X f \, \mathrm{d}\mu = \int_{x \in X} f(x) \, \mathrm{d}\mu(x).$$

Functions $f$ for which this integral is well-defined are called $\mu$-integrable. For example, every bounded measurable function $f: X \to \mathbb{R}_{\geq 0}$ is $\mu$-integrable. For $E \in \Sigma_X$, let

$$\int_E f \, \mathrm{d}\mu := \int_X \mathbb{1}_E \cdot f \, \mathrm{d}\mu.$$

**Proposition 2.9.** *Let $(X_1, \Sigma_{X_1})$ and $(X_2, \Sigma_{X_2})$ be measurable spaces, $\mu \in \mathbb{M}(\Sigma_{X_1})$ and $f: X_1 \to X_2$ a measurable function. The function $g: X_2 \to \mathbb{R}_{\geq 0}$ is $f\#\mu$-integrable iff $g \circ f$ is $\mu$-integrable. Further,*

$$\int_{X_2} g \, \mathrm{d}f\#\mu = \int_{X_1} g \circ f \, \mathrm{d}\mu.$$

**Combining measurable spaces**

**Definition 2.10** (Product $\sigma$-algebra). Let $(X_1, \Sigma_{X_1})$ and $(X_2, \Sigma_{X_2})$ be measurable spaces. The *product $\sigma$-algebra* $\Sigma_{X_1} \otimes \Sigma_{X_2}$ is the $\sigma$-algebra of $X_1 \times X_2$ defined by:

$$\Sigma_{X_1} \otimes \Sigma_{X_2} := \sigma(\{\, E_1 \times E_2 \mid E_i \in \Sigma_{X_i} \,\}).$$

For measures $\mu_i$ on $\Sigma_{X_i}$ let the product measure $\mu_1 \times \mu_2$ be the unique measure on $\Sigma_{X_1} \otimes \Sigma_{X_2}$ with $(\mu_1 \times \mu_2)(E_1 \times E_2) = \mu_1(E_1)\mu_2(E_2)$ for all $E_i \in \Sigma_{X_i}$.

The structure of $\mathbb{R}^n$, specifically that $\mathbb{R}$ is a Hausdorff space with countable base, allows us to decompose its Borel $\sigma$-algebra into a product of Borel $\sigma$-algebras on $\mathbb{R}$. This fact will be used frequently in our work, as it allows to consider a selection of the components of a vector in $\mathbb{R}^n$.

**Proposition 2.11.** $\mathcal{B}(\mathbb{R}^n) = \bigotimes_{i \in [1,n]} \mathcal{B}(\mathbb{R})$.

As it turns out, integrating with respect to a product measure corresponds to the iterated integral over the components in any order. This fundamental result of measure theory is known as Fubini's theorem:

$$\int_{X_1 \times X_2} f \, \mathrm{d}\mu_1 \times \mu_2 = \int_{X_i} \int_{X_j} f \, \mathrm{d}\mu_j \, \mathrm{d}\mu_i \text{ for } i \neq j \in \{1, 2\}.$$

Another common characterization of the product $\sigma$-algebra is that it is the smallest $\sigma$-algebra which makes the projection mappings

$$\mathrm{Proj}_{X_i} = \mathrm{Proj}_{X_1 \times X_2 \to X_i} : X_1 \times X_2 \to X_i, (x_1, x_2) \mapsto x_i$$

measurable. This gives rise to the notion of cylinder sets:

**Definition 2.12** (Cylinder set). Let $(X_1, \Sigma_{X_1})$ and $(X_2, \Sigma_{X_2})$ be measurable spaces. For $E_i \in \Sigma_{X_i}$ we define its *cylinder set* in $\Sigma_{X_1} \otimes \Sigma_{X_2}$ as the measurable set

$$\mathrm{Cyl}_{\Sigma_{X_1} \otimes \Sigma_{X_2}}(E_i) := \mathrm{Proj}_{X_1 \times X_2 \to X_i}^{-1}(E_i) \in \Sigma_{X_1} \otimes \Sigma_{X_2}.$$

We will simply write $\mathrm{Cyl}(E_i)$ if the product algebra is clear from the context.

**Proposition 2.13.** *Let $(X_1, \Sigma_{X_1})$ and $(X_2, \Sigma_{X_2})$ be measurable spaces.*

- *For $E, E' \in \Sigma_{X_i}$: $\mathrm{Cyl}(E) \cap \mathrm{Cyl}(E') = \mathrm{Cyl}(E \cap E')$.*
- *For $E_i \in \Sigma_{X_i}$: $\mathrm{Cyl}(E_1) \cap \mathrm{Cyl}(E_2) = \mathrm{Cyl}(E_1 \times E_2) =: \mathrm{Cyl}(E_1 E_2)$.*

**Proposition 2.14.** *Let $(X_1, \Sigma_{X_1})$, $(X_2, \Sigma_{X_2})$ and $(Y, \Sigma_Y)$ be measurable spaces. For functions $f_i \colon Y \to X_i, i \in [1, 2]$: $(f_1, f_2) \colon Y \to X_1 \times X_2$ is measurable iff both $f_1$ and $f_2$ are measurable.*

Another way of combining spaces is as a direct sum, an explicit disjoint union of the $\sigma$-algebras.

**Definition 2.15** (Direct sum $\sigma$-algebra). Let $(X_1, \Sigma_{X_1})$ and $(X_2, \Sigma_{X_2})$ be measurable spaces. The *direct sum $\sigma$-algebra* $\Sigma_{X_1} \oplus \Sigma_{X_2}$ is the $\sigma$-algebra of $(X_1 \times \{1\}) \,\dot\cup\, (X_2 \times \{2\})$ defined by:

$$\Sigma_{X_1} \oplus \Sigma_{X_2} := \sigma(\{(E_1 \times \{1\}) \,\dot\cup\, (E_2 \times \{2\}) \mid E_i \in \Sigma_{X_i}\}).$$

For measures $\mu_i$ on $\Sigma_{X_i}$ let the direct sum measure $\mu_1 \oplus \mu_2$ be the unique measure on $\Sigma_{X_1} \oplus \Sigma_{X_2}$ with $(\mu_1 \oplus \mu_2)((E_1 \times \{1\}) \,\dot\cup\, (E_2 \times \{2\})) = \mu_1(E_1) + \mu_2(E_2)$ for all $E_i \in \Sigma_{X_i}$.

## 2.3 Markov Models

Stochastic processes are used to model random behaviour as a sequence of (related) random variables. Of particular interest is the theory of Markov processes [69], a specific type of stochastic process in which the evolution of the process only depends on the current state. Mathematical formalizations of these processes in its most general form, often involve so-called (Markov or transition) kernels [28, 54].

**Definition 2.16** (Kernel). Let $(X_1, \Sigma_{X_1})$ and $(X_2, \Sigma_{X_2})$ be measurable spaces. A (finite) *kernel from* $\Sigma_{X_1}$ *to* $\Sigma_{X_2}$ is a function $\kappa \colon X_1 \times \Sigma_{X_2} \to \mathbb{R}_{\geq 0}$ such that

- $\kappa[x_1] \coloneqq \kappa(x_1, \cdot) \colon \Sigma_{X_2} \to \mathbb{R}_{\geq 0}$ is a measure for all $x_1 \in X_1$,
- $\kappa(\cdot, E_2) \colon X_1 \to \mathbb{R}_{\geq 0}$ is a $(\Sigma_{X_1}, \mathcal{B}(\mathbb{R}_{\geq 0}))$-measurable function for all $E_2 \in \Sigma_{X_2}$.

$\kappa$ is called a (sub-)probability kernel, if $\kappa[x_1]$ is a (sub-)probability measure for all $x_1 \in X_1$. We write $\kappa \colon X_1 \leadsto X_2$ to indicate that $\kappa$ is a kernel from $\Sigma_{X_1}$ to $\Sigma_{X_2}$.

Stated differently, a kernel associates a finite measure in $X_2$ with every element in $X_1$ in a measurable way. Thus, equivalently, through the process of currying, one can see a kernel as a measurable function $\kappa \colon X_1 \to \mathbb{M}(X_2)$. The $\sigma$-algebra on $\mathbb{M}(X_2)$ is the natural Giry $\sigma$-algebra on measures: $\sigma(\{\, \{\, \mu \in \mathbb{M}(X_2) \mid \mu(Q) \in B \,\} \mid Q \in \Sigma_{X_2}, B \in \mathcal{B}(\mathbb{R}_{\geq 0}) \,\})$.

A measure $\mu \colon \Sigma_X \to \mathbb{R}_{\geq 0}$ can be identified with the kernel $1 \leadsto X, ((), E) \mapsto \mu(E)$ from the singleton space $1 = \{\, ()\, \}$. We will use this correspondence and refer to this kernel as $\mu$.

We lift the notion of a property holding $\mu$-almost-surely to kernels: A property holds $\kappa$-almost-surely ($\kappa$-a.s.) if it holds $\kappa[x_1]$-a.s. for every $x_1 \in X_1$.

In the context of probabilistic timed automata, kernels are used in the definition of an automata's semantics for the representation of transitions between states. Intuitively, such a kernel provides, for every state and possible action, a distribution of states reached from the state by performing the action.

In this work we look at the more general setting of probabilistic timed automata with parametric probabilities. As a consequence, we shift from distributions to parametric distributions. Intuitively, such a parametric distribution consists of a concrete distribution over some space $X$ for every parameter valuation $\theta \in \Theta$. Conveniently, this can also be expressed using a kernel $\Theta \leadsto X$.

Kernels thus provide a proper measure-theoretic formalization of parametric distributions and transition probabilities. We define parametric kernels [53] to combine these two aspects into one object to represent parametric transition probabilities.

**Definition 2.17** (Parametric kernel). Given measurable spaces $(\Theta, \Sigma_\Theta)$, $(X_1, \Sigma_{X_1})$, and $(X_2, \Sigma_{X_2})$, a $\Theta$-*parametric kernel* (simply called *parametric kernel*) $\kappa \colon X_1 \leadsto_\Theta X_2$ is a kernel $\kappa \colon \Theta \times X_1 \leadsto X_2$, where $\Theta \times X_1$ is equipped with the corresponding product $\sigma$-algebra.

We lift the terminology regarding (sub-)probability kernels to parametric kernels. We sometimes write $\kappa[(\theta, \cdot)]$ as $\kappa_\theta[\cdot]$ for $\theta \in \Theta$ to highlight the distinction between parameter space and state space. Non-parametric kernels are subsumed by choosing a singleton set $\Theta = 1$ for the parameter space. Also, any non-parametric kernel can be turned into a parametric kernel by simply ignoring the input parameter value.

Parametric kernels can be composed in sequence. Intuitively, this corresponds to combining a sequence of transitions into a single kernel. Importantly, the parameter valuation is passed-through to every involved kernel.

**Definition 2.18** (Kernel composition). Let $(\Theta, \Sigma_\Theta)$ and $(X_i, \Sigma_{X_i})$ be measurable spaces, $i \in [1,3]$. Given parametric kernels $\kappa_1 \colon X_1 \rightsquigarrow_\Theta X_2$ and $\kappa_2 \colon X_2 \rightsquigarrow_\Theta X_3$, we define the $\Theta$-*pass-through composition* $\kappa_1 \overset{\Theta}{\fatsemi} \kappa_2 \colon X_1 \rightsquigarrow_\Theta X_3$ as a kernel with

$$(\kappa_1 \overset{\Theta}{\fatsemi} \kappa_2)[(\theta, x_1)](E_3) \coloneqq \int_{x_2 \in X_2} \kappa_2[(\theta, x_2)](E_3) \, \mathrm{d}\kappa_1[(\theta, x_1)](x_2),$$

for $\theta \in \Theta, x_1 \in X_1, E_3 \in \Sigma_{X_3}$, i.e. the valuation $\theta$ is used to evaluate both $\kappa_1$ and $\kappa_2$. For non-parametric kernels we simply write $\kappa_1 \fatsemi \kappa_2$ for $\kappa_1 \overset{1}{\fatsemi} \kappa_2$.

This pass-through of the parameter value for the evaluation of the kernels is very similar to a more general way to compose two kernels, known as the kernel product. It reuses the input for the first kernel together with its output to evaluate the second kernel, and as output also gives the value of this intermediate result. Again, parameter valuations are simply passed-through unaffected.

**Definition 2.19** (Kernel product). Let $(\Theta, \Sigma_\Theta)$ and $(X_i, \Sigma_{X_i})$ be measurable spaces, $i \in [1,3]$. Given kernels $\kappa_1 \colon X_1 \rightsquigarrow_\Theta X_2$ and $\kappa_2 \colon X_1 \times X_2 \rightsquigarrow_\Theta X_3$, the *product* $\kappa_1 \overset{\Theta}{\otimes} \kappa_2 \colon X_1 \rightsquigarrow_\Theta X_2 \times X_3$ is a kernel with

$$(\kappa_1 \overset{\Theta}{\otimes} \kappa_2)[(\theta, x_1)](E_{23}) \coloneqq \int_{x_2 \in X_2} \int_{x_3 \in X_3} \mathbb{1}_{E_{23}}(x_2, x_3) \, \mathrm{d}\kappa_2[(\theta, (x_1, x_2))](x_3) \, \mathrm{d}\kappa_1[(\theta, x_1)](x_2),$$

for $\theta \in \Theta, x_1 \in X_1, E_{23} \in \Sigma_{X_2} \otimes \Sigma_{X_3}$. For non-parametric kernels we simply write $\kappa_1 \otimes \kappa_2$ for $\kappa_1 \overset{1}{\otimes} \kappa_2$.

**Parametric Markov Decision Processes**

Markov Decision Processes (MDPs) are non-deterministic labelled transition systems in which the transitions are probabilistic. In the presence of uncountable state spaces, this behaviour is usually represented through Markov kernels, also called transition kernels. For every pair of state and available transition label, the kernel provides a distribution of reached states.

When considering parametric MDPs, the transition probabilities are usually parametrized through the use of rational functions over some indeterminates that describe the probability, i.e. measure, of a certain transition occurring [30, 50].

**Definition 2.20** (Polynomial, rational function). Let $X$ be a set of $|X| = n$ indeterminates. A *polynomial over* $\mathbb{Q}$ *in* $X$ is $\sum_{k \in \mathbb{N}_0^n} a_k X^k$ where $a_k \in \mathbb{Q}$ for all $k \in \mathbb{N}_0^n$, and $a_k \neq 0$ for only finitely-many $k$. The ring of all polynomials over $\mathbb{Q}$ in $X$ is denoted by $\mathbb{Q}[X]$. For a polynomial $f \in \mathbb{Q}[X]$ and a valuation $\sigma \in \mathrm{Val}(X) = \mathbb{Q}^X$, substitution of $x \in X$ by $\sigma(x)$ in $f$ yields $f[\sigma] \in \mathbb{Q}$. The field of fractions $\mathbb{Q}(X)$ consists of *rational functions* $\frac{p}{q}$, where $p, q \in \mathbb{Q}[X]$, $q \neq 0$.

For instance, the probability of transitioning between two states might be given as $p + \frac{1}{2}$, where $p$ is some indeterminate. When instantiating $p$ with a value from $[0, \frac{1}{2}]_\mathbb{Q}$, one obtains an actual probability. Applying this to every parametric expression in a Markov model thus yields the instantiated distributions. To formalize this using parametric kernels, we use the valid valuations of indeterminates as the parameter space of the kernel. The indeterminates themselves thus only occur implicitly in the defining rational function expressions.

**Definition 2.21** (Rational function parametric kernel). Given a set P of parameters and measurable spaces $(\Theta, \Sigma_\Theta)$, $(X_1, \Sigma_{X_1})$, $(X_2, \Sigma_{X_2})$. We say a parametric kernel $\kappa \colon X_1 \leadsto_\Theta X_2$ is a $\mathbb{Q}(P)$-*parametrized kernel*, if

(i) $\Theta \subseteq \mathrm{Val}(P)$, and

(ii) there is a family $(\mathfrak{f}_{x_1, E_2})_{x_1 \in X_1, E_2 \in \Sigma_{X_2}}$ in $\mathbb{Q}(P)$, such that $\kappa[(\theta, x_1)](E_2) = \mathfrak{f}_{x_1, E_2}[\theta]$ for all $\theta \in \Theta$, $x_1 \in X_1$ and $E_2 \in \Sigma_{X_2}$.

We write $\kappa \colon X_1 \leadsto_{\mathbb{Q}(P)} X_2$ to indicate that $\kappa$ is a $\mathbb{Q}(P)$-parametrized kernel.

A similar construction with $\mathbb{Q}[P]$ instead of $\mathbb{Q}(P)$ yields $\mathbb{Q}[P]$-parametric kernels.

Note that restricting the set of parameters (or their valuations in case of $\mathbb{Q}(P)$-parametric kernels) is always possible by restricting the kernel to the corresponding $\sigma$-subalgebra.

**Definition 2.22.** A *partial probability kernel* $\kappa \colon X_1 \leadsto X_2$ is a kernel $\kappa \colon X_1 \leadsto X_2$, such that for all $x_1 \in X_1$, either $\kappa[x_1]$ is a probability measure or $\kappa[x_1] = 0$.

We use partial probability kernels with the convention that the zero measure is assigned to represent the absence of certain transitions in some states.

**Definition 2.23** (parametric Markov Decision Process). Given a measurable space of parameters $(\Theta, \Sigma_\Theta)$, a $\Theta$-*parametric Markov Decision Process* ($\Theta$-*pMDP*) is a tuple $((S, \Sigma_S), (\mathrm{Act}, \Sigma_{\mathrm{Act}}), \mathrm{step})$, where:

- $(S, \Sigma_S)$ is a measurable space, S is called the set of states,
- $(\mathrm{Act}, \Sigma_{\mathrm{Act}})$ is a measurable space, Act is called the set of action labels,
- $\mathrm{step} \colon S \times \mathrm{Act} \leadsto_\Theta S$ is a $\Theta$-parametric partial probability kernel, called the transition function/relation.

We usually refer to a pMDP by the tuple $(S, \mathrm{Act}, \mathrm{step})$, leaving the associated parameter space $\Theta$ and $\sigma$-algebras implicit. A classical, non-parametric, Markov Decision Process (MDP) is a pMDP with $\Theta = 1$. A (parametric) Markov Chain (pMC) $(S, \mathrm{step})$ is an unlabelled (parametric) MDP, i.e. a pMDP with $\mathrm{Act} = 1$.

**Remark 2.24.** When specifying pMDPs we allow using (parametric) sub-probability distributions in step. This simplifies modelling of missing transitions. By introducing a distinguished, new sink state $\mathfrak{s}_\bot$ to which all the 'missing' probability mass leads, one can ensure that step is a (parametric) partial probability kernel.

**Schedulers**

Starting from an initial distribution of states, performing a sequence of transitions in a pMDP results in a parametric distribution of states. This distribution depends on the actions selected along the way. Such a choice of actions is done by a *scheduler*, also called strategy or adversary in the literature. Specifically, a scheduler induces a (parametric) distribution over the set of infinite paths.

**Definition 2.25** (Path). Given a pMDP $\mathcal{M} = ((S, \Sigma_S), (\text{Act}, \Sigma_{\text{Act}}), \text{step})$. A *path* of length $n \in \mathbb{N}_0$ is a finite alternating sequence of states and action labels, $\pi = (x_i)_{i \in [0, 2n]}$, with:

- $x_{2i} \in S$ for all $i \in [0, n]$, and
- $x_{2i-1} \in \text{Act}$ for all $i \in [1, n]$.

We write $|\pi| := n$ for the length, i.e. number of edges, of $\pi$. For $i \in [0, n]$, $\pi[i] := x_{2i}$ denotes the $i$-th state on the path and $\mathsf{last}(\pi) := \pi[|\pi|]$ refers to the last state on the path. For $i \in [1, n]$, $\text{act}_i(\pi) := x_{2i-1}$ denotes the path's $i$-th transition label. We will usually write a path as

$$\pi = \pi[0] \xrightarrow{\text{act}_1(\pi)} \pi[1] \xrightarrow{\text{act}_2(\pi)} \cdots \xrightarrow{\text{act}_n(\pi)} \pi[n].$$

The measurable space of paths of length $n \in \mathbb{N}_0$ is a product $\sigma$-algebra:

$$(\mathsf{Paths}_{\mathcal{M}}^n, \Sigma_{\mathsf{Paths}_{\mathcal{M}}^n}) = (S \times (\text{Act} \times S)^n, \Sigma_S \otimes (\Sigma_{\text{Act}} \otimes \Sigma_S)^n),$$

The measurable space of finite paths is the direct sum of the spaces of length $n \in \mathbb{N}_0$:

$$(\mathsf{Paths}_{\mathcal{M}}^{<\omega}, \Sigma_{\mathsf{Paths}_{\mathcal{M}}^{<\omega}}) = (\dot{\bigcup_{n \in \mathbb{N}_0}} \mathsf{Paths}_{\mathcal{M}}^n, \bigoplus_{n \in \mathbb{N}_0} \Sigma_{\mathsf{Paths}_{\mathcal{M}}^n}).$$

The measurable space of infinite paths whose $\sigma$-algebra is the $\sigma$-algebra generated by the cylinder sets of all finite paths is:

$$(\mathsf{Paths}_{\mathcal{M}}^{\omega}, \Sigma_{\mathsf{Paths}_{\mathcal{M}}^{\omega}}) = ((S \times \text{Act})^{\omega}, \Sigma_S \otimes (\Sigma_{\text{Act}} \otimes \Sigma_S)^{\omega}).$$

The measurable space of all, finite and infinite, paths is:

$$(\mathsf{Paths}_{\mathcal{M}}, \Sigma_{\mathsf{Paths}_{\mathcal{M}}}) = (\mathsf{Paths}_{\mathcal{M}}^{<\omega} \dot{\cup} \mathsf{Paths}_{\mathcal{M}}^{\omega}, \Sigma_{\mathsf{Paths}_{\mathcal{M}}^{<\omega}} \oplus \Sigma_{\mathsf{Paths}_{\mathcal{M}}^{\omega}}).$$

With a proper $\sigma$-algebra for states, we can now lift the one-step transition function step to paths so that for each path and action it gives a (parametric) distribution over the paths obtained by appending the transitions possible according to step to the given path.

**Definition 2.26.** Given a $\Theta$-pMDP $\mathcal{M} = ((S, \Sigma_S), (\text{Act}, \Sigma_{\text{Act}}), \text{step})$. The *path-lifted transition kernel* $\overline{\text{step}} \colon \mathsf{Paths}_{\mathcal{M}}^{<\omega} \times \text{Act} \rightsquigarrow_\Theta \mathsf{Paths}_{\mathcal{M}}^{<\omega}$ is defined as

$$\overline{\text{step}}_\theta[\pi, \alpha] := \mathsf{app}_{\pi,\alpha} \# \text{step}_\theta[\mathsf{last}(\pi), \alpha], \text{ for } \theta \in \Theta, \pi \in \mathsf{Paths}_{\mathcal{M}}^{<\omega}, \alpha \in \text{Act},$$

where $\mathsf{app}_{\pi,\alpha} \colon S \to \mathsf{Paths}_{\mathcal{M}}^{<\omega}, s \mapsto \pi \xrightarrow{\alpha} s$ is the measurable $(\pi, \alpha)$-section mapping.

A scheduler is now defined to take a finite path and give a distribution of transitions to take from the last state in the path. The path-lifted transition kernel is used in the definition of a scheduler to ensure that it only chooses *valid* transition labels, i.e. those which are not assigned the zero measure in the state they are selected in.

**Definition 2.27** (Scheduler). Given a $\Theta$-pMDP $\mathcal{M} = ((S, \Sigma_S), (\text{Act}, \Sigma_{\text{Act}}), \text{step})$. A *scheduler* $\sigma$ of $\mathcal{M}$ is a $\Theta$-parametric probability kernel $\sigma \colon \mathsf{Paths}_{\mathcal{M}}^{<\omega} \rightsquigarrow_\Theta \text{Act}$, such that $\sigma \overset{\Theta \times \mathsf{Paths}_{\mathcal{M}}^{<\omega}}{;} \overline{\text{step}} \colon \mathsf{Paths}_{\mathcal{M}}^{<\omega} \rightsquigarrow_\Theta \mathsf{Paths}_{\mathcal{M}}^{<\omega}$ is a $\Theta$-parametric probability kernel. The set of all schedulers on $\mathcal{M}$ is denoted by $\text{Sched}_{\mathcal{M}}$.

A scheduler is *deterministic*, if for all $\theta \in \Theta, \pi \in \mathsf{Paths}_{\mathcal{M}}^{<\omega}$: $\sigma_\theta[\pi] = \mathfrak{d}_{\alpha_{\theta,\pi}}$ for some $\alpha_{\theta,\pi} \in \mathrm{Act}$. A deterministic scheduler can also be seen as a measurable function $\Theta \times \mathsf{Paths}_{\mathcal{M}}^{<\omega} \to \mathrm{Act}$.

A scheduler is *memory-less*, if for $\pi, \pi' \in \mathsf{Paths}_{\mathcal{M}}^{<\omega}$: $last(\pi) = last(\pi') \implies \sigma[(\cdot, \pi)] = \sigma[(\cdot, \pi')]$, i.e. the schedulers choice depends only on the current, i.e. last, state of a path. Such a scheduler can also be viewed as a $\Theta$-parametric probability kernel $\mathrm{S} \rightsquigarrow_\Theta \mathrm{Act}$. The set of all memoryless, deterministic schedulers on $\mathcal{M}$ is denoted by $\mathrm{Sched}_{\mathcal{M}}^{\mathsf{MD}}$.

These subtypes of schedulers play an important role for Markov models. They are often easier to represent and reason about and, under certain assumptions, are sufficient to obtain optimal results for many objectives. For example, on finite MDPs it is sufficient to consider memoryless, deterministic schedulers to maximize the probability of reaching a set of target states [69].

A scheduler completely determines the behaviour of a pMDP. Due to the restriction that a scheduler must always be able to choose a next action, no deadlocked-states can be reached. As a consequence, every finite path can be extended infinitely and we can restrict our attention to infinite paths to define the pMDP's behaviour. The distribution of infinite paths in the pMDP is obtained by choosing successors to extend all finite paths according to the scheduler.

**Definition 2.28.** A $\Theta$-pMDP $\mathcal{M} = ((\mathrm{S}, \Sigma_\mathrm{S}), (\mathrm{Act}, \Sigma_\mathrm{Act}), \mathsf{step})$, a probability measure $\iota \colon 1 \rightsquigarrow \mathrm{S}$ and a scheduler $\sigma \colon \mathsf{Paths}_{\mathcal{M}}^{<\omega} \rightsquigarrow_\Theta \mathrm{Act}$ induce a unique $\Theta$-*parametric probability distribution on* $\mathsf{Paths}_{\mathcal{M}}^{\omega}$, denoted $\mathbb{P}\mathrm{r}_{\iota,\sigma}^{\mathcal{M}} \colon 1 \rightsquigarrow_\Theta \mathsf{Paths}_{\mathcal{M}}^{\omega}$ with:

$$\mathbb{P}\mathrm{r}_{\iota,\sigma}^{\mathcal{M}}[\theta](\mathrm{Cyl}(S_0 \times \bigtimes_{i \in [1,n]} (A_i \times S_i))) = (\iota \overset{\Theta}{\otimes} \bigotimes_{i \in [1,n]}^{\Theta} (\sigma \overset{\Theta}{\otimes} \overline{\mathsf{step}}))(S_0 \times \bigtimes_{i \in [1,n]} (A_i \times S_i)),$$

for $\theta \in \Theta$, $n \in \mathbb{N}_0$, $S_i \in \Sigma_\mathrm{S}$ for $i \in [0, n]$, $A_i \in \Sigma_\mathrm{Act}$ for $i \in [1, n]$.

We write $\mathbb{P}\mathrm{r}_{s,\sigma}^{\mathcal{M}} := \mathbb{P}\mathrm{r}_{\mathfrak{d}_s,\sigma}^{\mathcal{M}}$ for the path probability distribution starting from $s \in \mathrm{S}$.

Observe that the initial distribution $\iota \colon 1 \rightsquigarrow \mathrm{S}$ is non-parametric, thus giving the same initial state for every parameter valuation. Note that by Carathéodory's extension theorem a measure defined as above is only uniquely determined if it is $\sigma$-finite. But this is always ensured, as we restrict our attention to (parametric) probability measures for pMDPs.

**Reachability Probabilities**

A common problem in the analysis of MDPs is the optimization, i.e. minimization or maximization, of reachability probabilities for a certain set of target states [37].

**Definition 2.29** (Reachability target). Given a $\Theta$-pMDP $\mathcal{M} = ((\mathrm{S}, \Sigma_\mathrm{S}), (\mathrm{Act}, \Sigma_\mathrm{Act}), \mathsf{step})$ and measurable set $T \in \Sigma_\mathrm{S}$. An infinite path $\pi \in \mathsf{Paths}_{\mathcal{M}}^{\omega}$ reaches $T$, written $\pi \vDash \Diamond T$, if $\pi[i] \in T$ for some $i \in \mathbb{N}_0$. The set of *all infinite paths that reach* $T$ is

$$[\![\Diamond T]\!] = \{ \pi \in \mathsf{Paths}_{\mathcal{M}}^{\omega} \mid \pi \vDash \Diamond T \}$$
$$= \dot{\bigcup}_{k \in \mathbb{N}_0} \mathrm{Cyl}(((\mathrm{S} \setminus T) \times \mathrm{Act})^k \times T).$$

In particular, $[\![\Diamond T]\!]$ is a countable union of measurable sets and thus measurable. When no confusion arises, we will simply write $\Diamond T$ for $[\![\Diamond T]\!]$.

In the parametric setting this extends to the problem of obtaining optimal parameter valuations. For example in safety applications one tries to decrease the probability of a fault in the worst case, i.e. one tries to find parameter valuations that minimize the maximal probability of an error occurring.

**Definition 2.30** (Reachability probabilities). Let $\mathcal{M} = ((S, \Sigma_S), (\text{Act}, \Sigma_{\text{Act}}), \text{step})$ be a $\Theta$-pMDP, $\iota\colon 1 \rightsquigarrow S$ an initial state probability distribution and $T \in \Sigma_S$ a measurable target set. The *maximal/minimal reachability probability of $T$ from $\iota$ in $\mathcal{M}$* is

$$\overline{\mathbb{Pr}}_\iota^\mathcal{M}[\theta](\lozenge T) := \sup_{\sigma \in \text{Sched}_\mathcal{M}} \mathbb{Pr}_{\iota,\sigma}^\mathcal{M}[\theta](\lozenge T),$$

$$\underline{\mathbb{Pr}}_\iota^\mathcal{M}[\theta](\lozenge T) := \inf_{\sigma \in \text{Sched}_\mathcal{M}} \mathbb{Pr}_{\iota,\sigma}^\mathcal{M}[\theta](\lozenge T).$$

For a set of parameter valuations $\Theta' \subseteq \Theta$ we let

$$\overline{\mathbb{Pr}}_{\iota,\max\Theta'}^\mathcal{M}(\lozenge T) := \sup_{\theta \in \Theta'} \overline{\mathbb{Pr}}_\iota^\mathcal{M}[\theta](\lozenge T),$$

$$\overline{\mathbb{Pr}}_{\iota,\min\Theta'}^\mathcal{M}(\lozenge T) := \inf_{\theta \in \Theta'} \overline{\mathbb{Pr}}_\iota^\mathcal{M}[\theta](\lozenge T),$$

and $\underline{\mathbb{Pr}}_{\iota,\max\Theta'}^\mathcal{M}(\lozenge T)$ and $\underline{\mathbb{Pr}}_{\iota,\min\Theta'}^\mathcal{M}(\lozenge T)$ defined analogously.

Decision and synthesis problems for reachability in parametric MDPs often involve different quantifications over the schedulers [52, 50]. For example, for $\mathcal{Q} \in \{\exists, \forall\}$, $\Theta' \subseteq \Theta$, $\lambda \in [0,1]_\mathbb{R}$ and $\bowtie \in \{<, \leq, =, \geq, >\}$, one can consider the problem

$$\exists \theta \in \Theta' \colon \mathcal{Q}\sigma \in \text{Sched}_\mathcal{M} \colon \mathbb{Pr}_{\iota,\sigma}^\mathcal{M}[\theta](\lozenge T) \bowtie \lambda,$$

i.e., searching for a valuation of parameters such that the reachability probability of a target fulfills a certain threshold ($\bowtie \lambda$) for some scheduler ($\mathcal{Q} = \exists$) or for all possible schedulers ($\mathcal{Q} = \forall$).

If all infima/suprema are attained, i.e. are actual minima/maxima, the notation defined above allows to abstract the quantifiers and enables the representation of such problems using a single inequality. As an example, consider the safety application of trying to limit the probability of a fault occurring in the worst case. Looking for such a parameter valuation, corresponds to the problem of checking whether there exists a parameter valuation such that for all schedulers the reachability probability is at most $\lambda$, i.e.

$$\exists \theta \in \Theta' \colon \forall \sigma \in \text{Sched}_\mathcal{M} \colon \mathbb{Pr}_{\iota,\sigma}^\mathcal{M}[\theta](\lozenge T) \leq \lambda.$$

That all schedulers fulfill a certain probability upper bound $\lambda$ is equivalent to the supremum of the probabilities fulfilling this upper bound, i.e.

$$\exists \theta \in \Theta' \colon \overline{\mathbb{Pr}}_\iota^\mathcal{M}[\theta](\lozenge T) \leq \lambda.$$

And this in turn is fulfilled, iff the inequality

$$\overline{\mathbb{Pr}}_{\iota,\min\Theta'}^\mathcal{M}(\lozenge T) \leq \lambda$$

holds.

Parameter space partitioning is a common approach to tackle this reachability synthesis problem [51]. More generally, given a property in some logic, the goal is to partition a region of parameter valuations into sets of fulfilling and falsifying valuations. As a complete categorization is often infeasible, approximating approaches that try to classify a certain percentage of the total parameter space are used instead.

# 3 Parameters in Probabilistic Timed Automata

In this section we will formally define probabilistic timed automata with parameters in both transition probabilities and clock constraints. The clocks in timed automata always evolve at the same speed and cannot be stopped. The only operation affecting the clocks is a reset to zero, possible when performing a transition. To restrict the possible behaviour in timed automata, one can use transition guards and location invariants. These are specified through constraints whose syntax and semantics is specified before we can give the formal definition of the probability-parametric clock-parametric probabilistic timed automata, later in this section.

## 3.1 Clock Constraints

Timed automata have a finite set of clocks which evolve at a common constant rate. The clocks can be reset to zero when taking a transition. These transitions are commonly guarded using certain linear constraints on the clock values. Imagining a grid of possible clock valuations, the constraints allow to restrict the clock valuations to rectangular regions, hence the name rectangular guards. In clock-parametric timed automata [45], parameters for these clock constraints were introduced. More specifically, instead of comparing clocks against fixed constants, a comparison with parameters is allowed whose value is determined through a valuation.

**Definition 3.1** (Clock constraints). Given a finite set Clk of clocks and a finite set P of parameters, the set of *parametrized clock constraints* $\text{CONSTR}_P(\text{Clk})$ is the smallest set generated by the following grammar:

$$\varphi := c_1 \bowtie t \mid c_1 - c_2 \bowtie t \mid \varphi \wedge \varphi,$$

where $c_1, c_2 \in \text{Clk}$, $t \in \mathbb{N}_0 \mathbin{\dot{\cup}} P$ and $\bowtie \in \{\leq, \geq, =, <, >\}$.

When $P = \emptyset$, we write $\text{CONSTR}(\text{Clk}) = \text{CONSTR}_P(\text{Clk})$ for the set of (non-parametric) clock constraints.

For a partial parameter valuation $\gamma \colon P' \to \mathbb{N}_0, P' \subseteq P$, we let $\varphi[\gamma] \in \text{CONSTR}_{P \setminus P'}(\text{Clk})$ denote the constraint obtained from $\varphi$ by replacing parameters $p \in P'$ by their value $\gamma(p)$.

To evaluate a parametric clock constraint, one needs a valuation for the clocks as well as for the parameters. We aggregate them in a single combined valuation.

**Definition 3.2** ((Combined) valuation). Given a finite set Clk of clocks and a finite set P of parameters. The set of *(combined) valuations* $\text{Val}(\text{Clk}, P)$ consists of pairs $(\tau, \gamma)$, with $\tau \colon \text{Clk} \to \mathbb{R}_{\geq 0}$, and $\gamma \colon P \to \mathbb{N}_0$. We will sometimes identify $(\tau, \gamma)$ with the combined function $\tau \mathbin{\dot{\cup}} \gamma \colon \text{Clk} \mathbin{\dot{\cup}} P \to \mathbb{R}_{\geq 0} \cup \mathbb{N}_0$.

**Definition 3.3.** Let $\nu = \tau \dot{\cup} \gamma \in \mathrm{Val}(\mathrm{Clk}, \mathrm{P})$ be a valuation over clocks Clk and parameters P, and $\varphi \in \mathrm{Constr}_\mathrm{P}(\mathrm{Clk})$ be a parametric clock constraint. We say $\nu$ *fulfils* $\varphi$, written $\nu \vDash \varphi$ or $\tau \vDash \varphi[\gamma]$, defined inductively on the structure of the formula:

- $\nu \vDash c_1 \bowtie t \iff \nu(c_1) \bowtie \nu(t)$,
- $\nu \vDash c_1 - c_2 \bowtie t \iff \nu(c_1) - \nu(c_2) \bowtie \nu(t)$,
- $\nu \vDash \varphi_1 \wedge \varphi_2 \iff \nu \vDash \varphi_1$ and $\nu \vDash \varphi_2$.

The set of valuations fulfilling $\varphi$ is denoted by $[\![\varphi]\!] \coloneqq \{\, \nu \in \mathrm{Val}(\mathrm{Clk}, \mathrm{P}) \mid \nu \vDash \varphi \,\}$.

Note that there is no negation operation in the language of constraints. On an atomic level negation is possible through the use of a suitable dual for $\bowtie$. But importantly there is no disjunction, which results in the satisfying valuations being convex, rectangular, sets.

**Proposition 3.4.** *For every $\varphi \in \mathrm{Constr}_\mathrm{P}(\mathrm{Clk})$, the set $[\![\varphi]\!] \subseteq \mathbb{R}_{\geq 0}^{|\mathrm{Clk}|} \times \mathbb{N}_0^{|\mathrm{P}|}$ is convex and measurable.*

Observe that, since the sets of clocks and parameters are finite, there are only countably-many different constraints. This allows us to use the power set $\mathcal{P}\left(\mathrm{Constr}_\mathrm{P}(\mathrm{Clk})\right)$ as the standard $\sigma$-algebra for $\mathrm{Constr}_\mathrm{P}(\mathrm{Clk})$, hence every singleton $\{\,\varphi\,\}$ is measurable.

Also note that, when there are multiple subformulas of the same form present in a constraint, the *strongest* one suffices, e.g. $c_1 < 2 \wedge c_1 < 5$ is semantically equivalent to $c_1 < 2$, that is $[\![c_1 < 2 \wedge c_1 < 5]\!] = [\![c_1 < 2]\!]$. This is related to the canonicalization operation for parametric Difference Bounded Matrices [2, 45] which are a common representation of constraints and symbolic states in timed automata. From now on we will consider all constraints up to equivalence of $[\![\cdot]\!]$.

Before we use the constraints to define the pPpTA, we quickly recap the definition of common operations on clock valuations [24]. These operations include letting time elapse and resetting clocks. For combined valuations these are defined to only affect the clocks' values and leave the parameter valuation unchanged.

**Definition 3.5.** Given a combined valuation $\nu = \tau \dot{\cup} \gamma \in \mathrm{Val}(\mathrm{Clk}, \mathrm{P})$.

The valuation $\nu + \delta$ obtained by *delaying* for $\delta \in \mathbb{R}_{\geq 0}$ time is defined as

$$(\nu + \delta)(x) \coloneqq \begin{cases} \tau(x) + \delta, & \text{if } x \in \mathrm{Clk}, \\ \gamma(x), & \text{if } x \in \mathrm{P}. \end{cases}$$

The valuation $\nu[R]$ obtained by *resetting clocks* in $R \subseteq \mathrm{Clk}$ is defined as

$$\nu[R](x) \coloneqq \begin{cases} 0, & \text{if } x \in R, \\ \nu(x), & \text{else.} \end{cases}$$

For set of valuations $Val \subseteq \mathrm{Val}(\mathrm{Clk}, \mathrm{P}_{\mathrm{Clk}})$ and set of clocks $R \subseteq \mathrm{Clk}$ we define

- the valuations obtained by *clock reset*: $Val[R] \coloneqq \{\, \nu[R] \mid \nu \in Val \,\}$,
- the valuations obtained by *inverse clock reset*: $Val[R^{-1}] \coloneqq \{\, \nu \in \mathrm{Val}(\mathrm{Clk}, \mathrm{P}_{\mathrm{Clk}}) \mid \nu[R] \in Val \,\}$.

For set of valuations $Val, Val' \subseteq \text{Val}(\text{Clk}, \text{P}_{\text{Clk}})$ we define

- the *time successors* of $Val$ always staying in $Val'$:

$$\nearrow_{Val'}(Val) := \left\{ \nu' \in \text{Val}(\text{Clk}, \text{P}_{\text{Clk}}) \;\middle|\; \begin{array}{l} \exists \nu \in Val, \delta \in \mathbb{R}_{\geq 0} : \nu' = \nu + \delta, \text{ and} \\ \forall \delta' \in [0, \delta]_\mathbb{R} : \nu + \delta' \in Val' \end{array} \right\},$$

- the *time predecessors* of $Val$ always staying in $Val'$:

$$\swarrow_{Val'}(Val) := \left\{ \nu' \in \text{Val}(\text{Clk}, \text{P}_{\text{Clk}}) \;\middle|\; \begin{array}{l} \exists \nu \in Val, \delta \in \mathbb{R}_{\geq 0} : \nu' + \delta = \nu, \text{ and} \\ \forall \delta' \in [0, \delta]_\mathbb{R} : \nu' + \delta' \in Val' \end{array} \right\}.$$

## 3.2 Timed Automata with Unknown Probabilities and Clock Constraints

We can now give a formal definition of probability-parametric clock-parametric probabilistic timed automata. We separate parameters used for clock constraints from parameters used to determine the probability distributions of transitions. The clock-parametric aspect as known from classical parametric timed automata [45] is represented using the introduced parametric constraints. The parametrization of transition probabilities is represented using a parametric transition kernel. To be in-line with existing approaches to probability-parametric MDPs [50] and probability-parametric probabilistic timed automata [40], we restrict to *rational function parametrizations*.

**Definition 3.6** (p$\mathbb{P}$pTA)**.** A *probability-parametric, clock-parametric Probabilistic Timed Automaton* (*p$\mathbb{P}$pTA*) is a tuple $\mathcal{P} = (\text{Loc}, \text{Act}, \text{Clk}, \text{P}, \text{inv}, \text{guard}, \text{trans}, l_0)$ where:

- Loc is a finite set of *locations*,
- Act is a finite set of *actions* (*labels*),
- Clk is a finite set of *clocks*,
- $\text{P} = \text{P}_{\text{Clk}} \,\dot{\cup}\, \text{P}_{prob}$ is a finite, disjoint set of *parameters for clocks* $\text{P}_{\text{Clk}}$ and *parameters for probabilities* $\text{P}_{prob}$,
- $\text{inv} \colon \text{Loc} \to \text{Constr}_{\text{P}_{\text{Clk}}}(\text{Clk})$ assigns to every location a parametric clock constraint, the *location invariant*,
- $\text{guard} \colon \text{Loc} \times \text{Act} \to \text{Constr}_{\text{P}_{\text{Clk}}}(\text{Clk})$ assigns to every potential transition a parametric clock constraint, the *transition guard*,
- $\text{trans} \colon \text{Loc} \times \text{Act} \rightsquigarrow_{\mathbb{Q}(\text{P}_{prob})} \mathcal{P}(\text{Clk}) \times \text{Loc}$ the $\mathbb{Q}(\text{P}_{prob})$-parametric partial probability kernel, called the *(Act-labelled) transition function*,
- $l_0 \in \text{Loc}$ is the *initial location*.

We write $\text{Val}_\mathcal{P}(\text{P}_{prob}) \subseteq \text{Val}(\text{P}_{prob})$ for the set of valid probability parameter valuations that is implicitly defined through the parameter space of the parametric kernel trans.

Note that as Loc, Act, Clk and P are all finite sets, we use the corresponding power set as their associated $\sigma$-algebra. We can thus define all the functions using the appropriate measure-theoretic constructs, even though this might seem superfluous due to countability of the involved sets. Using this common notation makes it straight-forward to formally define the pPpTAs' semantics, where we inevitably obtain uncountable state spaces due to the presence of continuous time.

**Definition 3.7** (Instantiation). Let $\mathcal{P}$ be a pPpTA. For $P' \subseteq P$ with $P'_{prob} = P' \cap P_{prob}$, $P'_{Clk} = P' \cap P_{Clk}$, a partial parameter valuation is a mapping $\gamma \colon P' \to \mathbb{Q}$ with $\gamma(P'_{Clk}) \subseteq \mathbb{N}_0$. $\gamma$ is called (full) parameter valuation, if $P' = P$.

We obtain the *(partially) instantiated pPpTA* $\mathcal{P}[\gamma] = (\text{Loc}, \text{Act}, \text{Clk}, (P_{Clk} \setminus P'_{Clk}) \dot\cup (P_{prob} \setminus P'_{prob}), \text{inv}', \text{guard}', \text{trans}', l_0)$ by substituting all occurrences of $p \in P'$ in $\mathcal{P}$ by $\gamma(p)$ where

- $\text{inv}' \colon \text{Loc} \to \text{Constr}_{P_{Clk} \setminus P'_{Clk}}(\text{Clk}), l \mapsto \text{inv}(l)[\gamma \restriction_{P'_{Clk}}]$,
- $\text{guard}' \colon \text{Loc} \times \text{Act} \to \text{Constr}_{P_{Clk} \setminus P'_{Clk}}(\text{Clk}), (l, \alpha) \mapsto \text{guard}(l, \alpha)[\gamma \restriction_{P'_{Clk}}]$,
- $\text{trans}' \colon \text{Loc} \times \text{Act} \rightsquigarrow_{\mathbb{Q}(P_{prob} \setminus P'_{prob})} \mathcal{P}(\text{Clk}) \times \text{Loc}, (\bar{\gamma}, l, \alpha, \cdot) \mapsto \text{trans}(\gamma \dot\cup \bar{\gamma}, l, \alpha, \cdot)$.

Our definition of pPpTAs generalizes and includes existing models from the literature.

**Remark 3.8.** Given a pPpTA $\mathcal{P}$ with parameters $P = P_{Clk} \dot\cup P_{prob}$.

- If $P_{Clk} = \emptyset = P_{prob}$, $\mathcal{P}$ is a Probabilistic Timed Automaton (PTA) as defined in [62].
- If $P_{Clk} = \emptyset \neq P_{prob}$, $\mathcal{P}$ is a probability-parametric Probabilistic Timed Automaton (pPTA) as defined in [40].
- If $P_{Clk} \neq \emptyset = P_{prob}$, $\mathcal{P}$ is a clock-parametric Probabilistic Timed Automaton (PpTA) as defined in [47].
- If $P_{prob} = \emptyset$ and trans only maps to Dirac distributions or the zero measure, i.e. trans is a (partial) measurable function $\text{Loc} \times \text{Act} \rightharpoonup \text{Loc}$, then $\mathcal{P}$ is a clock-parametric Timed Automaton (pTA) [3]; if additionally $P_{Clk} = \emptyset$, $\mathcal{P}$ is a Timed Automaton (TA) [2].

**Example 3.9.** Consider a variation of the probabilistic non-repudiation protocol in [67] which was modelled using PTA in [68]. Herein, one participant, the originator, tries to establish common knowledge with a second party, the recipient. The recipient is untrusted and could quit participation in the exchange at any point. The goal is to exchange the information, without one-sidedly giving information away in case of repudiation, i.e. when either party denies having participated in all or part of the exchange.

The protocol [67] achieves this through a series of message exchanges, where the total number of messages to be exchanged is determined by a geometric distribution. The originator needs to send these messages to the recipient one-by-one, waiting for an acknowledgment before continuing. The channel used by the originator to transmit the message is faulty and can drop a message with some unknown probability. After successfully sending the message to the sender, the originator waits for an acknowledgment which is assumed to be transmitted flawlessly. Further, the originator uses a timer to restrict the time it waits for another action to occur. If the timer expires without any action occurring, the computation is also considered to have failed.

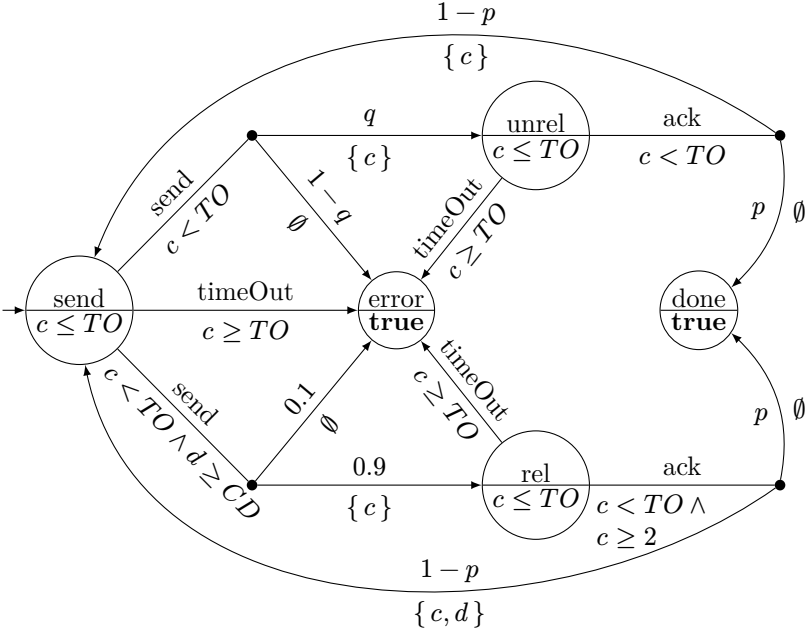

Figure 1: A pℙpTA modelling a parametrized variation of the probabilistic non-repudiation protocol of [67].

This behaviour is modelled as part of the pℙpTA depicted in Figure 1, via the locations *send* and *unrel*. The probability of successfully transmitting the message to the recipient is modelled with a *probability parameter* $q$ that can take values in $(0,1)_\mathbb{R}$. A clock $c$ is used to track that progress occurs timely, i.e. before a time-out which is specified by a *clock parameter TO*. After every successful acknowledgment, the protocol successfully completes with *(parametric) probability* $p$ or continues with another message to be sent. This effectively models a geometric distribution in $p$ of the number of messages exchanged in the communication.

We extend the scenario by another option for the originator to transmit the message to the recipient through a different channel, e.g. a middle man. This other option is known to have a reliability of 90% when delivering the message from originator to recipient, however the acknowledgments from the recipient take at least 2 seconds to be delivered. Additionally, this alternative transport channel is not always available. After every use it has a certain waiting period, during which this option is not available. In the pℙpTA this is modelled through the alternative path from *send* to *done* via *rel*. The cooldown delay is modelled using an additional *clock parameter CD* and an additional clock $d$ is used to keep track of when the second channel becomes available again.

# 4 Semantics and Reachability Problems

To complete the definition of pℙpTA we will provide their semantics and consider the formal definition of various notions of reachability in this section.

## 4.1 Semantics

The semantics of a pℙpTA is an uncountably large *parametric Markov Decision Process*. Each state in this pMDP combines a location in the pℙpTA with a valuation of current clock values and clock parameter values. In every state, one can either let time elapse deterministically as long as this is allowed by the current location's invariant; or a discrete labelled action from the timed automaton can be taken, if the guard is fulfilled. When performing any of these transitions, the location and clock values can change, but the parameter valuation remains unchanged. Thus, the semantics can be seen as a combination of copies of the semantics for pℙTA by [40], one for each valid parameter valuation. This is analogous to the definition of [45] for pTA semantics based on TA semantics.

**Definition 4.1** (Concrete semantics). The *concrete semantics of a pℙpTA* $\mathcal{P} = (\text{Loc}, \text{Act}, \text{Clk}, \text{P}, \text{inv}, \text{guard}, \text{trans}, l_0)$ is the $\mathbb{Q}(\text{P}_{prob})$-parametric MDP $[\![\mathcal{P}]\!] := (\mathbb{S}, \text{Act} \,\dot\cup\, \mathbb{R}_{\geq 0}, step)$ where:

- $\mathbb{S} := \{(l, \nu) \in \text{Loc} \times \text{Val}(\text{Clk}, \text{P}_{\text{Clk}}) \mid \nu \vDash \text{inv}(l)\}$ is the set of states, with the $\sigma$-algebra $\Sigma_{\mathbb{S}}$ being the product $\sigma$-algebra using the embedding of $\text{Val}(\text{Clk}, \text{P}_{\text{Clk}}) \subseteq \mathbb{R}_{\geq 0}^{|\text{Clk}|} \times \mathbb{N}_0^{|\text{P}_{\text{Clk}}|}$,

- action labels $\text{Act} \,\dot\cup\, \mathbb{R}_{\geq 0}$ using, with some abuse of notation, the direct sum algebra $\mathcal{P}(\text{Act}) \oplus \mathcal{B}(\mathbb{R}_{\geq 0})$,

- $step \colon \mathbb{S} \times (\text{Act} \,\dot\cup\, \mathbb{R}_{\geq 0}) \rightsquigarrow_{\mathbb{Q}(\text{P}_{prob})} \mathbb{S}$ the $\mathbb{Q}(\text{P}_{prob})$-parametric sub-probability kernel, defined for $(l, \nu) \in \mathbb{S}, \rho \in \text{Val}_{\mathcal{P}}(\text{P}_{prob})$ by:

    - For $\alpha \in \text{Act}$:
    $$step_\rho[(l, \nu), \alpha] := \mathbb{1}_{[\![\text{guard}(l,\alpha)]\!]}(\nu) \cdot at_\nu \# \text{trans}_\rho[l, \alpha],$$
    where $at_\nu \colon \mathcal{P}(\text{Clk}) \times \text{Loc} \to \mathbb{S}, (R, l') \mapsto (l', \nu[R])$ is a measurable function that applies the effect of taking the transition, i.e. moving to location $l'$ and resetting clocks in $R$, in the given valuation $\nu$.

    - For $\delta \in \mathbb{R}_{\geq 0}$:
    $$step_\rho[(l, \nu), \delta] := \mathbb{1}_{[\![\text{inv}(l)]\!]}(\nu + \delta) \cdot \mathfrak{d}_{(l, \nu + \delta)},$$
    which applies the effect of delaying for $\delta$ time.

We write $\text{Val}_{\mathcal{P}}(\text{P}_{\text{Clk}}) := \{\gamma \in \text{Val}(\text{P}_{\text{Clk}}) \mid (l_0, \underline{0} \,\dot\cup\, \gamma) \in \mathbb{S}\}$ for the set of valid clock parameter valuations, i.e. those valuations for which the state in the initial location and with all clock values set to zero is valid. For every $\gamma \in \text{Val}_{\mathcal{P}}(\text{P}_{\text{Clk}})$, we define the initial distribution under valuation $\gamma$ as the Dirac probability distribution $\iota_\gamma := \mathfrak{d}_{(l_0, \underline{0} \,\dot\cup\, \gamma)} \colon 1 \rightsquigarrow \mathbb{S}$ in the initial location with all clocks set to zero.

The definition of a pℙpTA's semantics, especially $at_\nu$, contains the implicit assumption that no invalid states outside $\mathbb{S}$, i.e. with $\nu \not\models \text{inv}(l)$, can be reached from a state in $\mathbb{S}$ when performing transitions. This is commonly referred to as a well-formedness assumption for timed automata and can easily be ensured by propagating location invariants backwards along transitions [1, 24].

Also note that in the definition of time delay transitions for timed automata it is usually ensured that not only the final state at $\nu + \delta$ fulfils the location invariant at $l$, but that this is the case for every state $\nu + \delta'$ with $\delta' \in [0, \delta]_\mathbb{R}$. In our definition, it suffices to check $\nu + \delta \models \text{inv}(l)$ by the convexity of $[\![\text{inv}(l)]\!]$ (Proposition 3.4) and the assumption $\nu \models \text{inv}(l)$.

The concept of schedulers introduced in Section 2 is directly applicable to the semantics of pℙpTA. Choosing Act-labelled transitions which are disabled by a guard or performing time delays which would violate the location invariants is prevented directly by the definition of schedulers.

To realistically model the behaviour of timed automata, special care needs to be taken of the transitions representing time delays. While the Act-labelled transitions should be interpreted as instant internal state changes of the automata, the time delays represent actual evolution of time. By choosing ever smaller time steps $\delta$, or even $\delta = 0$, a scheduler could construct an infinite run which would only take a finite amount of time. Such executions are known as Zeno-runs [2]. As we strive to represent realistically achievable behaviour, the usual restriction of schedulers which only produce non-Zeno paths is necessary.

**Definition 4.2** (Divergent path). Given a pℙpTA $\mathcal{P}$. For a path $\pi \in \text{Paths}_{[\![\mathcal{P}]\!]}^\omega$ in its concrete semantics and $k \in \mathbb{N}_0$, let $\text{dur}(\pi, k) := \sum_{\substack{i \in [1,k] \\ \text{act}_i(\pi) \in \mathbb{R}_{\geq 0}}} \text{act}_i(\pi)$ denote the duration of path $\pi$ in the first $k$ steps. $\pi$ is called *divergent*, iff $\lim_{k \to \infty} \text{dur}(\pi, k) = \infty$.

Observe that the set of all divergent paths is measurable by a straight-forward generalization of the argument in [61] for ℙTAs.

**Definition 4.3** (Non-Zeno scheduler). For a pℙpTA $\mathcal{P}$, a scheduler $\sigma \colon \text{Paths}_{[\![\mathcal{P}]\!]}^{<\omega} \rightsquigarrow_{\mathbb{Q}(P_{prob})} \text{Act} \;\dot{\cup}\; \mathbb{R}_{\geq 0}$ on $[\![\mathcal{P}]\!]$ is *almost-surely diverging/non-Zeno* if for all initial probability distributions $\iota_\gamma, \gamma \in \text{Val}_\mathcal{P}(P_{\text{Clk}})$: path $\pi \in \text{Paths}_{[\![\mathcal{P}]\!]}^\omega$ is divergent $\mathbb{Pr}_{\iota, \sigma}^{[\![\mathcal{P}]\!]}$-almost-surely.

With some abuse of notation we will sometimes refer to a scheduler of $[\![\mathcal{P}]\!]$ simply as a scheduler of $\mathcal{P}$. Before continuing, we make some assumptions to simplify the presentation in the remainder of this work.

**Assumption 4.4.** For the remainder of this work:

- Constraints of pℙpTAs, in both invariants and transition guards, are assumed to include no diagonal constraints, i.e. there are no constraints of the form $c_1 - c_2 \bowtie t$. For TAs a simple construction [18] eliminates diagonal constraints, but incurs an exponential increase of the model size, proportional to the number of constraints to eliminate. This construction can be analogously applied to pℙpTAs.

- We restrict our attention to almost-surely diverging schedulers. That is, any reference to schedulers of $[\![\mathcal{P}]\!]$ is meant to refer to almost-surely diverging schedulers.

## 4.2 Reachability Problems

We now turn our attention to the analysis of properties of pPpTA. In the context of model checking with parametric models, parameter synthesis plays a crucial role. In the remainder of this work we want to explore possibilities to synthesize optimal parameter valuations for pPpTAs, focussing on reachability objectives.

Reachability objectives represent an important subset of properties that are more generally expressed using a logic, usually real/dense-time CTL (called TCTL) or in its probabilistic version PTCTL [38]. Classically, for many formulae model checking can be reduced to different reachability properties [44, 64, 12]. The important difference to classical reachability objectives $\Diamond F$ is that, due to the dense time progression, reachability formulae can now be satisfied *in-between* transitions if the transition represents the passage of time. Intuitively, the goal is also reached if it is just passed through.

**Definition 4.5** (Dense reachability target). Given a pPpTA $\mathcal{P}$ and its concrete semantics $[\![\mathcal{P}]\!] = (\mathbb{S}, \mathrm{Act} \dot\cup \mathbb{R}_{\geq 0}, step)$. A set $F \subseteq \mathrm{Loc} \times \mathrm{Val}(\mathrm{Clk})$ such that $\hat{F} = (F \times \mathrm{Val}(\mathrm{P}_{\mathrm{Clk}})) \cap \mathbb{S} \in \Sigma_{\mathbb{S}}$ is measurable, is called *target* or *goal*. A path $\pi \in \mathsf{Paths}^\omega_{[\![\mathcal{P}]\!]}$ *reaches* $F$, written $\pi \vDash \blacklozenge F$, iff for some $i \in \mathbb{N}_0$:

- $\pi[i] \in \hat{F}$, or
- $\mathrm{act}_i(\pi) = \delta \in \mathbb{R}_{\geq 0}$ and there exists $\delta' \in (0, \delta)_{\mathbb{R}}$ with $\pi[i-1] + \delta' \in \hat{F}$.

Similarly, the set of *all infinite paths that reach $F$* is denoted as $[\![\blacklozenge F]\!]$, or simply $\blacklozenge F$ when no confusion can arise. A construction similar to the one for $[\![\Diamond T]\!]$ shows that $[\![\blacklozenge F]\!]$ is a countable union of measurable sets and thus measurable [61, 28].

We will also simply write $L$ for the target $L \times \mathrm{Val}(\mathrm{Clk})$ where $L \subseteq \mathrm{Loc}$, or even just $l$, in case $L = \{\, l \,\}$.

Note that a target set can only refer to and put constraints on the clocks of the timed automaton. Valuations of the clock-parameters are separated from the goal. Often, reachability targets are given as pairs $(l, \varphi)$ of locations $l \in \mathrm{Loc}$ and parameter-free constraints $\varphi \in \mathrm{CONSTR}(\mathrm{Clk})$, describing a set $[\![\varphi]\!]$ of valid clock valuations.

With this separation of parameters from target states, one can now consider the problem of finding optimal parameter values for dense-time reachability objectives in pPpTAs, similar to the reachability problems for pMDPs.

**Definition 4.6.** For pPpTA $\mathcal{P}$ and $\gamma \in \mathrm{Val}_\mathcal{P}(\mathrm{P}_{\mathrm{Clk}})$ we write $\overline{\mathbb{Pr}}^\mathcal{P}_\gamma = \overline{\mathbb{Pr}}^{[\![\mathcal{P}]\!]}_{\iota_\gamma}$ and $\underline{\mathbb{Pr}}^\mathcal{P}_\gamma = \underline{\mathbb{Pr}}^{[\![\mathcal{P}]\!]}_{\iota_\gamma}$. Recall that in both definitions the extrema are taken over all almost-surely diverging schedulers, cf. Assumption 4.4. Further, we define for $\Theta_{\mathrm{Clk}} \subseteq \mathrm{Val}_\mathcal{P}(\mathrm{P}_{\mathrm{Clk}}), \Theta_{prob} \subseteq \mathrm{Val}_\mathcal{P}(\mathrm{P}_{prob})$:

$$\overline{\mathbb{Pr}}^{\mathcal{P}}_{\gamma, \max \Theta_{prob}}(\blacklozenge F) := \sup_{\rho \in \Theta_{prob}} \overline{\mathbb{Pr}}^{\mathcal{P}}_\gamma[\rho](\blacklozenge F),$$

$$\overline{\mathbb{Pr}}^{\mathcal{P}}_{\max \Theta_{\mathrm{Clk}}, \Theta_{prob}}(\blacklozenge F) := \sup_{\gamma \in \Theta_{\mathrm{Clk}}} \overline{\mathbb{Pr}}^{\mathcal{P}}_{\gamma, \max \Theta_{prob}}(\blacklozenge F),$$

$$\overline{\mathbb{Pr}}^{\mathcal{P}}_{\gamma, \min \Theta_{prob}}(\blacklozenge F) := \inf_{\rho \in \Theta_{prob}} \overline{\mathbb{Pr}}^{\mathcal{P}}_\gamma[\rho](\blacklozenge F),$$

$$\overline{\mathbb{Pr}}^{\mathcal{P}}_{\min \Theta_{\mathrm{Clk}}, \Theta_{prob}}(\blacklozenge F) := \inf_{\gamma \in \Theta_{\mathrm{Clk}}} \overline{\mathbb{Pr}}^{\mathcal{P}}_{\gamma, \min \Theta_{prob}}(\blacklozenge F),$$

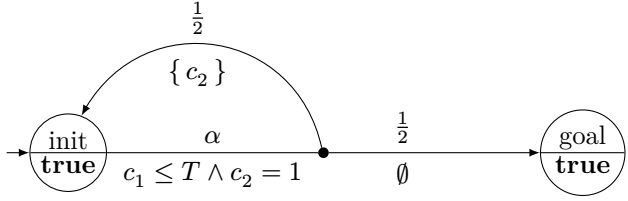

Figure 2: A $\mathbb{P}$pTA representing a geometric loop whose number of iterations is bound by the clock parameter $T$.

and $\underline{\mathbb{Pr}}^{\mathcal{P}}_{\gamma,\max\Theta_{prob}}(\blacklozenge F)$, $\underline{\mathbb{Pr}}^{\mathcal{P}}_{\max\Theta_{\text{Clk}},\Theta_{prob}}(\blacklozenge F)$, $\underline{\mathbb{Pr}}^{\mathcal{P}}_{\gamma,\min\Theta_{prob}}(\blacklozenge F)$, and $\underline{\mathbb{Pr}}^{\mathcal{P}}_{\min\Theta_{\text{Clk}},\Theta_{prob}}(\blacklozenge F)$ defined analogously.

For example, this allows us to formally specify queries searching for optimal parameter valuations that minimize the maximal probability of some failure state occurring within a certain time bound. However, note that while the parameters for clocks and probabilities are separated, we only consider optimizing both parameters in the *same direction*, i.e. minimizing/maximizing. Although it is possible to give more general definitions, we have refrained from doing so here as it would further complicate matters.

Ideally, one would like to obtain optimal parameter valuations, i.e. ones which really give the greatest/least possible probability, turning the infima/suprema into minima/maxima. Unfortunately, the combination of probabilistic transitions together with the dense time progression makes this not possible in general, as the following example shows.

**Example 4.7.** Consider the $\mathbb{P}$pTA $\mathcal{P}$ (i.e. without parametric probabilities) depicted in Figure 2. Given a valuation $\gamma \in \text{Val}_{\mathcal{P}}(\text{P}_{\text{Clk}})$ with $\gamma(T) \geq 1$, the $\mathbb{P}$TA $\mathcal{P}[\gamma]$ admits the following behaviour: After delaying for exactly one time unit, the transition $\alpha$ can be taken, resulting in reaching the goal location with probability $\frac{1}{2}$ or returning to the initial location from where it can wait again for one time unit to repeat action $\alpha$. This loop can be repeated up to $x := \lfloor \gamma(T) \rfloor$ times, since clock $c_1$ is never reset. This results in

$$\overline{\mathbb{Pr}}^{\mathcal{P}}_{\gamma}(\blacklozenge\text{goal}) = 1 - \frac{1}{2^x}.$$

In particular, there always remains some non-zero probability of not reaching the target. It is clear to see that this probability is always increasing towards 1 as $x \to \infty$, i.e. $\gamma(T) \to \infty$, but never reaches it. Thus, with $\Theta_{\text{Clk}} = \mathbb{N}_0$ and $\Theta_{prob} = \{()\}$ we get

$$\overline{\mathbb{Pr}}^{\mathcal{P}}_{\max\Theta_{\text{Clk}},\Theta_{prob}}(\blacklozenge\text{goal}) = 1.$$

But this supremum is not obtained by any $\gamma \in \Theta_{\text{Clk}}$. One would need to consider $\mathbb{P}$TA $\mathcal{P}[\gamma_\infty]$ with the improper valuation $\gamma_\infty$ that assigns $\infty$ to $T$, thus effectively dropping the constraint $c_1 \leq T$ from transition $\alpha$. This makes an unbounded number of loop traversals possible for a scheduler, hence results in achieving the optimal value, i.e. $\gamma_\infty$ is an optimizing, but improper, valuation.

As can be seen in the example, in the presence of probabilistic transitions, the clock parameters introduce the possibility of no proper valuation achieving an optimal reachability probability. As common in such situations [51, 40], we restrict our attention to closed and bounded regions of parameter valuations to ensure the existence of optimal solutions. It will often be helpful to further restrict to so-called rectangular regions.

**Definition 4.8** (Rectangular region). Let $\mathcal{P}$ be a pℙpTA and $\Theta_{\text{Clk}} \subseteq \text{Val}_{\mathcal{P}}(\text{P}_{\text{Clk}})$. $\Theta_{\text{Clk}}$ is *rectangular in* $p \in \text{P}_{\text{Clk}}$, if $\{p\}$ is separable in $\Theta_{\text{Clk}}$ and $\Theta_{\text{Clk}}\!\restriction_p = [l_p, u_p]$ for some $l_p, u_p \in \mathbb{N}_0$. $\Theta_{\text{Clk}}$ is rectangular if it is rectangular in every $p \in \text{P}_{\text{Clk}}$, i.e. it is of the form $\Theta_{\text{Clk}} = \bigtimes_{p \in \text{P}_{\text{Clk}}} [l_p, u_p]$ for some $l_p, u_p \in \mathbb{N}_0$ for each $p \in \text{P}_{\text{Clk}}$.

# 5 Lower and Upper Bound Parameters

Already in the classical case of timed automata, deciding whether some target states are reachable is undecidable in the presence of clock parameters [45]. Our more general formalism thus inherits all these complications while also adding new challenges. In the classical setting, this motivated the search for fragments of clock-parametric timed automata on which such reachability objectives become decidable. One approach is to determine certain syntactic criteria that restrict the set of definable automata in a suitable manner.

The most prominent example are L/U-automata [45]. These are pTAs in which each parameter can only ever be used as either a lower or upper bound in all constraints of the automaton. It was briefly mentioned in [47] that this criterion could be adapted to ℙpTAs. In this section we will show that a similar extension of L/U-automata is possible for pℙpTAs and consequently allows to find optimizing valuations for certain clock parameters.

As mentioned above, the main concept in L/U-automata [45] is the categorization of clock parameters into lower and upper bound parameters.

**Definition 5.1** (L/U-automaton). Given a pℙpTA $\mathcal{P}$. A clock parameter $p \in \text{P}_{\text{Clk}}$ is called a *lower bound parameter*, if for all $c \in \text{Clk}, \lesssim \in \{\leq, <, =\}$ no expression $c \lesssim p$ occurs in any constraint $\varphi \in \text{Im}(\text{inv}) \cup \text{Im}(\text{guard})$. Similarly, $p$ is an *upper bound parameter*, if for all $c \in \text{Clk}, \gtrsim \in \{\geq, >, =\}$ no $c \gtrsim p$ occurs in any $\varphi \in \text{Im}(\text{inv}) \cup \text{Im}(\text{guard})$.

$\mathcal{P}$ is called an *L-pℙpTA* if all $p \in \text{P}_{\text{Clk}}$ are lower bound parameters. Similarly, $\mathcal{P}$ is a *U-pℙpTA* if all $p \in \text{P}_{\text{Clk}}$ are upper bound parameters. $\mathcal{P}$ is an *L/U-pℙpTA*, if the clock parameters can be partitioned into $\text{P}_{\text{Clk}} = L \dot\cup U$ with lower bound parameters $L$ and upper bound parameters $U$.

For example, the pℙpTA depicted in Figure 1 has a lower bound parameter $CD$, but is not an L/U-pℙpTA because $TO$ occurs as both a lower and upper bound. The pℙpTA shown in Figure 2 is a U-pℙpTA.

For clock-parametric timed automata, the important observation for L/U-automata was that these satisfy a monotonicity property in the parameter valuations with respect to enabled transitions. Whenever a transition is possible for some parameter valuation, increasing upper bound parameters resp. decreasing lower bound parameters keeps this transition enabled, as this only relaxes the constraints. This property enables a reduction of the reachability problem from pTAs to classical TAs by choosing the least restrictive parameter valuation and checking whether the goal is reachable for this instantiation.

In the classical case, this least-restrictive (improper) valuation sets all lower parameters to 0 and all upper parameters to $\infty$, essentially making all constraints containing these parameters superfluous as they become trivially fulfilled.

**Definition 5.2.** An *extended clock parameter valuation* $\gamma \colon \mathrm{P}_{\mathrm{Clk}} \to \mathbb{N}_0 \,\dot\cup\, \{\infty\}$ allows assigning $\infty$ to parameters. An extended valuation $(\tau, \gamma)$ then consists of a valuation for clocks $\tau \colon \mathrm{Clk} \to \mathbb{R}_{\geq 0}$ (as before) and an extended clock parameter valuation $\gamma$. For the evaluation of constraints we assume $x < \infty$ for all $x \in \mathbb{R}$, as usual.

One thus considers the L/U-pTA instantiated with the extended parameter valuation $(0)_{p \in L} \,\dot\cup\, (\infty)_{p \in U}$. If the goal is not reachable therein, the aforementioned monotonicity property lets us conclude that there exists no viable proper, i.e. non-extended, valuation that reaches the target. The crucial observation is that, in case the target is reachable under this extended valuation, there actually exists a proper valuation which also allows to reach the goal. This is due to the fact that a path reaching the goal does so in finitely many steps. Consequently, all clock values occurring in this initial path fragment until reaching the goal can be bounded from above by some constant. Choosing any values above this threshold for all relevant upper bound parameters, so as to not invalidate any guards or invariants, thus results in proper valuations that can reach the target.

Unfortunately, this approach relies on the reachability question being qualitative, i.e. the question *whether* the goal is reachable or not. It is not immediately applicable to the probabilistic case, as the focus is shifted to a quantitative analysis, i.e. *how likely* it is to reach the goal. The number of ways to reach a goal becomes relevant. In general, such an extended evaluation can allow for a higher probability of reaching a goal than what any non-extended valuation could achieve.

**Example 5.3.** Reconsider the $\mathbb{P}$pTA $\mathcal{P}$ from Example 4.7. Indeed, the only clock parameter $T$ is an upper bound parameter, thus $\mathcal{P}$ is a U-$\mathbb{P}$pTA. We already saw that for $\gamma_1, \gamma_2 \in \mathrm{Val}_{\mathcal{P}}(\mathrm{P}_{\mathrm{Clk}})$ with $\gamma_1(T) \leq \gamma_2(T)$ we have

$$\overline{\mathbb{P}\mathrm{r}}_{\gamma_1}^{\mathcal{P}}(\blacklozenge \mathrm{goal}) \leq \overline{\mathbb{P}\mathrm{r}}_{\gamma_2}^{\mathcal{P}}(\blacklozenge \mathrm{goal}) < 1 = \overline{\mathbb{P}\mathrm{r}}_{\gamma_\infty}^{\mathcal{P}}(\blacklozenge \mathrm{goal}),$$

with extended parameter valuation $\gamma_\infty = (T \mapsto \infty)$. This is because $\gamma_\infty$ allows for an unbounded number of $\alpha$-transitions in $\mathcal{P}[\gamma_\infty]$, while this number is always bounded for ordinary valuations. Consequently, the maximal reachability probability of location goal under $\gamma_\infty$ is unattainable by proper valuations.

While this example illustrates the need to restrict to closed and bounded clock parameter regions to achieve any fruitful insights even for L/U-p$\mathbb{P}$pTAs, it also indicates that the mentioned monotonicity property extends to this more general setting involving (parametric) probabilities. Indeed, increasing U-parameters or decreasing L-parameters increases the maximal and decreases the minimal reachability probabilities.

**Proposition 5.4.** *Given a p$\mathbb{P}$pTA $\mathcal{P}$, a parameter valuation $\gamma \in \mathrm{Val}_{\mathcal{P}}(\mathrm{P}_{\mathrm{Clk}})$ and $p \in \mathrm{P}_{\mathrm{Clk}}$.*

- *If $p$ is a lower bound parameter: For all $\gamma' = \gamma[p \mapsto x] \in \mathrm{Val}_{\mathcal{P}}(\mathrm{P}_{\mathrm{Clk}})$ with $x \leq \gamma(p)$, every scheduler of $\mathcal{P}[\gamma]$ is also a scheduler of $\mathcal{P}[\gamma']$.*

  *In particular, for all targets $F \subseteq \mathrm{Loc} \times \mathrm{Val}(\mathrm{Clk})$ and $\rho \in \mathrm{Val}_{\mathcal{P}}(\mathrm{P}_{prob})$:*

  $$\overline{\mathbb{P}\mathrm{r}}_\gamma^{\mathcal{P}}[\rho](\blacklozenge F) \leq \overline{\mathbb{P}\mathrm{r}}_{\gamma'}^{\mathcal{P}}[\rho](\blacklozenge F) \qquad and \qquad \underline{\mathbb{P}\mathrm{r}}_\gamma^{\mathcal{P}}[\rho](\blacklozenge F) \geq \underline{\mathbb{P}\mathrm{r}}_{\gamma'}^{\mathcal{P}}[\rho](\blacklozenge F).$$

- *If $p$ is an upper bound parameter: For all $\gamma' = \gamma[p \mapsto x] \in \mathrm{Val}_{\mathcal{P}}(\mathrm{P}_{\mathrm{Clk}})$ with $x \geq \gamma(p)$, every scheduler of $\mathcal{P}[\gamma]$ is also a scheduler of $\mathcal{P}[\gamma']$.*

*In particular, for all targets $F \subseteq \text{Loc} \times \text{Val(Clk)}$ and $\rho \in \text{Val}_{\mathcal{P}}(P_{prob})$:*

$$\overline{\mathbb{Pr}}^{\mathcal{P}}_{\gamma}[\rho](\blacklozenge F) \leq \overline{\mathbb{Pr}}^{\mathcal{P}}_{\gamma'}[\rho](\blacklozenge F) \qquad and \qquad \underline{\mathbb{Pr}}^{\mathcal{P}}_{\gamma}[\rho](\blacklozenge F) \geq \underline{\mathbb{Pr}}^{\mathcal{P}}_{\gamma'}[\rho](\blacklozenge F).$$

To establish these results one only needs to show that schedulers of $\mathcal{P}[\gamma]$ are also schedulers of $\mathcal{P}[\gamma']$. But this is immediate, since the restriction that a valid scheduler under valuation $\gamma$ needs to almost-surely pick enabled actions is obviously also fulfilled when the scheduler is interpreted under valuation $\gamma'$. The only ways actions become enabled/disabled in pPpTAs is through transition guards and location invariants which prevent time delay actions. Both cases are realized with parametric rectangular constraints. Modifying the value of parameter $p$ in $\gamma'$ only relaxes any constraints involving $p$, while keeping all other constraints unchanged. Thus, whenever a constraint is fulfilled with $\gamma$, thus enabling the action in $[\![\mathcal{P}[\gamma]]\!]$, the guard will also be fulfilled by the relaxation $\gamma'$, i.e. the action is also enabled in $[\![\mathcal{P}[\gamma']]\!]$.

As a consequence, optimal valuations can be determined for lower and upper bound parameters when they are separable, thus reducing the number of parameters that need to be considered for parameter synthesis.

**Lemma 5.5.** *Given a pPpTA $\mathcal{P}$ and target $F \subseteq \text{Loc} \times \text{Val(Clk)}$. Let $\Theta_{\text{Clk}} \subseteq \text{Val}_{\mathcal{P}}(P_{\text{Clk}})$, $\Theta_{prob} \subseteq \text{Val}_{\mathcal{P}}(P_{prob})$ and $p \in P_{\text{Clk}}$, such that $p$ is separable in $\Theta_{\text{Clk}}$.*

*i) If $p$ is a lower bound parameter:*

$$\overline{\mathbb{Pr}}^{\mathcal{P}}_{\max \Theta_{\text{Clk}}, \Theta_{prob}}(\blacklozenge F) = \overline{\mathbb{Pr}}^{\mathcal{P}[p \mapsto \inf \Theta_{\text{Clk}} \upharpoonright_p]}_{\max \Theta_{\text{Clk}} \upharpoonright_{P_{\text{Clk}} \setminus \{p\}}, \Theta_{prob}}(\blacklozenge F)$$

*and*

$$\underline{\mathbb{Pr}}^{\mathcal{P}}_{\min \Theta_{\text{Clk}}, \Theta_{prob}}(\blacklozenge F) = \underline{\mathbb{Pr}}^{\mathcal{P}[p \mapsto \inf \Theta_{\text{Clk}} \upharpoonright_p]}_{\min \Theta_{\text{Clk}} \upharpoonright_{P_{\text{Clk}} \setminus \{p\}}, \Theta_{prob}}(\blacklozenge F).$$

*ii) If $p$ is an upper bound parameter:*

$$\overline{\mathbb{Pr}}^{\mathcal{P}}_{\max \Theta_{\text{Clk}}, \Theta_{prob}}(\blacklozenge F) = \overline{\mathbb{Pr}}^{\mathcal{P}[p \mapsto \sup \Theta_{\text{Clk}} \upharpoonright_p]}_{\max \Theta_{\text{Clk}} \upharpoonright_{P_{\text{Clk}} \setminus \{p\}}, \Theta_{prob}}(\blacklozenge F)$$

*and*

$$\underline{\mathbb{Pr}}^{\mathcal{P}}_{\min \Theta_{\text{Clk}}, \Theta_{prob}}(\blacklozenge F) = \underline{\mathbb{Pr}}^{\mathcal{P}[p \mapsto \sup \Theta_{\text{Clk}} \upharpoonright_p]}_{\min \Theta_{\text{Clk}} \upharpoonright_{P_{\text{Clk}} \setminus \{p\}}, \Theta_{prob}}(\blacklozenge F).$$

Intuitively, for lower bound parameters minimal and maximal reachability probabilities are attained by setting them to their *minimal* value, while for upper bound parameters this is achieved by using their *maximal* value.

In particular, when the parameter regions are closed and bounded, e.g. in rectangular regions, the suprema/infima are actually attained, thus we really get optimal valuations for these parameters and not just theoretically optimal values. Note that the requirement that the parameter is separable in the parameter valuation region is crucial, as demonstrated in the following example.

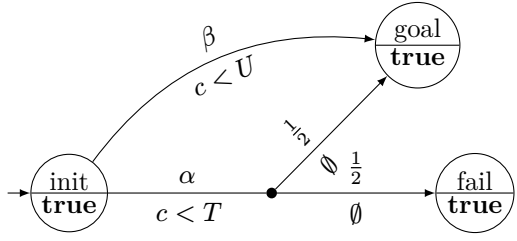

Figure 3: A ℙpTA showing the need for separability of parameter valuations.

**Example 5.6.** Consider the ℙpTA $\mathcal{P}$ depicted in Figure 3 with a single clock $c$ and clock parameter region $\Theta_{\text{Clk}} = \{(T \mapsto 1, U \mapsto 0), (T \mapsto 0, U \mapsto 1)\}$ (and $\Theta_{prob} = \{()\}$). Both $T$ and $U$ are upper bound parameters, but they are both not separable in $\Theta_{\text{Clk}}$.

Trying to optimize $\overline{\mathbb{P}r}^{\mathcal{P}}_{\max \Theta_{\text{Clk}},\Theta_{prob}}(\blacklozenge\text{goal})$ by looking at the upper bound parameters individually and first maximizing the value assigned to $T$ only leaves the valid valuation $\gamma = (T \mapsto 1, U \mapsto 0)$ remaining. But $\overline{\mathbb{P}r}^{\mathcal{P}}_{\gamma}(\blacklozenge\text{goal}) = \frac{1}{2}$, whereas the optimal valuation $\gamma_{opt} = (T \mapsto 0, U \mapsto 1)$ has $\overline{\mathbb{P}r}^{\mathcal{P}}_{\gamma_{opt}}(\blacklozenge\text{goal}) = 1$.

**Example 5.7.** Recall the modified probabilistic non-repudiation protocol from Example 3.9 which was modelled as a pℙpTA in Figure 1 (page 19). The model contains two clock parameters: $P_{\text{Clk}} = \{TO, CD\}$. Inspection of the model reveals that $CD$ is a lower bound parameter, while $TO$ is neither an upper nor lower bound parameter. Assume we are interested in determining parameter values that maximize the probability of reaching the target *done*. Further, assume that we consider values in $[6, 10]$ for $CD$ and $[3, 20]$ for $TO$, i.e. the clock parameter region $\Theta_{\text{Clk}}$ is rectangular. We can thus apply Lemma 5.5 to eliminate $CD$ and equivalently search for the maximal reachability probability in $\mathcal{P}[CD \mapsto \inf[6,10]] = \mathcal{P}[CD \mapsto 6]$ which only contains a single clock parameter:

$$\overline{\mathbb{P}r}^{\mathcal{P}}_{\max \Theta_{\text{Clk}},\Theta_{prob}}(\blacklozenge\text{done}) = \overline{\mathbb{P}r}^{\mathcal{P}[CD \mapsto 6]}_{\max (TO \in [3,20]),\Theta_{prob}}(\blacklozenge\text{done}).$$

We can slightly modify the model by not only considering time-outs but also cover other potential sources of error. One way to model this is to modify $\mathcal{P}$ by removing all constraints $c \geq TO$ from the transitions labelled with timeOut, effectively turning them into error-labelled transitions. Due to the location invariants, the non-Zenoness constraint on schedulers still ensures that the timeOut-transition is performed at the latest when a time-out occurs. In this changed pℙpTA $\tilde{\mathcal{P}}$, $TO$ is an upper bound parameter. With the same parameter region as above, we can thus use Lemma 5.5 to obtain

$$\overline{\mathbb{P}r}^{\tilde{\mathcal{P}}}_{\max \Theta_{\text{Clk}},\Theta_{prob}}(\blacklozenge\text{done}) = \overline{\mathbb{P}r}^{\tilde{\mathcal{P}}[CD \mapsto 6, TO \mapsto 20]}_{\max \Theta_{prob}}(\blacklozenge\text{done}).$$

In particular, $\tilde{\mathcal{P}}[CD \mapsto 6, TO \mapsto 20]$ is a pℙTA. We have thus completely eliminated clock parameters from the model and simplified the problem of determining parameter valuations that maximize the probability of reaching done in the pℙpTA $\tilde{\mathcal{P}}$ to a probability-parameter synthesis problem on a pℙTA.

**Theorem 5.8.** *Let $\mathcal{P}$ be an L-p$\mathbb{P}$pTA, U-p$\mathbb{P}$pTA or L/U-p$\mathbb{P}$pTA. For target $F \subseteq \mathrm{Loc} \times \mathrm{Val}(\mathrm{Clk})$ and rectangular $\Theta_{\mathrm{Clk}} \subseteq \mathrm{Val}_{\mathcal{P}}(\mathrm{P}_{\mathrm{Clk}})$ there exists a p$\mathbb{P}$TA $\hat{\mathcal{P}}$ with the same maximal/minimal reachability probability as $\mathcal{P}$, i.e. for $\Theta_{prob} \subseteq \mathrm{Val}_{\mathcal{P}}(\mathrm{P}_{prob})$:*

$$\overline{\mathbb{Pr}}^{\mathcal{P}}_{\max \Theta_{\mathrm{Clk}}, \Theta_{prob}}(\blacklozenge F) = \overline{\mathbb{Pr}}^{\hat{\mathcal{P}}}_{\max \Theta_{prob}}(\blacklozenge F) \quad and \quad \underline{\mathbb{Pr}}^{\mathcal{P}}_{\min \Theta_{\mathrm{Clk}}, \Theta_{prob}}(\blacklozenge F) = \underline{\mathbb{Pr}}^{\hat{\mathcal{P}}}_{\min \Theta_{prob}}(\blacklozenge F).$$

This allows us to reduce optimal parameter synthesis problems from L/U-p$\mathbb{P}$pTAs to the simpler model of p$\mathbb{P}$TAs, that only contain probability parameters. For these models, a variety of methods for reachability objectives is presented in [40]. Additionally, future results on these models can also be directly applied for the analysis of L/U-p$\mathbb{P}$pTAs.

# 6 Abstracting Semantics

The dense evolution of time in the semantics of timed automata induces an uncountably large state space, which makes enumeration of all states to analyze reachability infeasible. This is further amplified by the presence of parameters. If syntactical criteria, like L/U-parameters, are not applicable, other approaches like state space abstractions, which aim to reduce the size of the state space that needs to be analyzed, might aid in the analysis [24].

In this section, we consider an adaptation of the *backwards exploration semantics* [64] which was successfully extended to probability-parametric probabilistic timed automata in [40]. We also present a generalization of *digital clock semantics* [43, 61], which replaces the dense progress of time by integer time steps, to p$\mathbb{P}$pTAs. Together with suitable bounds on maximal clock values and parameter values, this then results in a finite abstraction for many relevant cases.

## 6.1 Symbolic Transitions

By abstracting the uncountable state space one tries to obtain a smaller, in the best case finite, representation of the original automaton's semantics, or an approximation thereof [14]. Symbolic semantics are based on such an abstraction of the concrete state space in the form of symbolic states. The goal is to obtain an (over)approximation of the concrete semantics by combining different clock valuations at the same location such that similar transitions are possible on these aggregated states.

Parameters integrate nicely into this approach, as parameter valuations can also be aggregated in the symbolic states.

**Definition 6.1** (Symbolic state, zone). Given a p$\mathbb{P}$pTA $\mathcal{P}$ and its concrete semantics $[\![\mathcal{P}]\!] = (\mathbb{S}, \mathrm{Act}\, \dot{\cup}\, \mathbb{R}_{\geq 0}, step)$. We define a *symbolic state* as a pair $(l, \zeta)$ with $l \in \mathrm{Loc}$ and $\zeta \subseteq \mathrm{Val}(\mathrm{Clk}, \mathrm{P}_{\mathrm{Clk}})$, such that $(\{l\} \times \zeta) \cap \mathbb{S} \in \Sigma_{\mathbb{S}}$ is measurable. We write $(l, \nu) \in (l, \zeta)$ for $(l, \nu) \in (\{l\} \times \zeta) \cap \mathbb{S}$.

A *zone* is a symbolic state $(l, [\![\varphi]\!])$ for some $\varphi \in \mathrm{CONSTR}_{\mathrm{P}_{\mathrm{Clk}}}(\mathrm{Clk})$.

Below we define operations on symbolic states that are commonly used [24, 45]. These lift the (inverse of) transitions on concrete states in a p$\mathbb{P}$pTA to the symbolic states by applying the corresponding operation individually on each contained concrete state.

Contrary to the clock valuations, parameter valuations are not modified by transitions and stay constant during an execution in a p$\mathbb{P}$pTA's concrete semantics. Hence, when performing a transition of the timed automaton on a symbolic state to potential next states, we can only shrink the set of possible parameter valuations, which happens whenever a parameter valuation restricts the transition guard in such a way that no available clock valuation fulfils the guard. Such concrete states are simply discarded from the symbolic state along a transition, which is similar to what happens in symbolic semantics with standard (non-parametric) clock valuations that do not fulfil the guard.

**Definition 6.2** (Symbolic time transitions). For pPpTA $\mathcal{P}$ and symbolic state $(l, \zeta)$ define

- the time delay successor symbolic state:

$$\mathsf{succ}_{time}(l, \zeta) := (l, \nearrow_{[\![\mathrm{inv}(l)]\!]}(\zeta)),$$

- the time delay predecessor symbolic state:

$$\mathsf{pre}_{time}(l, \zeta) := (l, \swarrow_{[\![\mathrm{inv}(l)]\!]}(\zeta)).$$

**Definition 6.3** (Symbolic discrete transitions). For pPpTA $\mathcal{P}$, symbolic state $(l, \zeta)$ and $l' \in \mathrm{Loc}, \alpha \in \mathrm{Act}, R \subseteq \mathrm{Clk}$ define

- the discrete successor for concrete transition $\bar{e} = (l, \alpha, R, l')$:

$$\mathsf{succ}_{\bar{e}}(l, \zeta) := (l', (\zeta \cap [\![\mathrm{guard}(l, \alpha)]\!])[R] \cap [\![\mathrm{inv}(l')]\!]),$$

- the discrete predecessor for concrete transition $\bar{e} = (l', \alpha, R, l)$:

$$\mathsf{pre}_{\bar{e}}(l, \zeta) := (l', \zeta[R^{-1}] \cap [\![\mathrm{guard}(l', \alpha)]\!] \cap [\![\mathrm{inv}(l')]\!]).$$

Observe that all those operations preserve zones [45]. When applied to a zone, the result can again be a represented as a zone. This can be easily seen by interpreting intersections as conjunctions and the (inverse) clock resets as appropriate relaxation or introduction of additional constraints, as done in the representation as Difference Bounded Matrices [31].

## 6.2 Backwards Exploration

The main idea of the backwards semantics of [64] for $\mathbb{P}$TAs is to start at the target of a reachability problem, represented as a symbolic state. From there one successively applies the predecessor operations defined above for all possible transitions to explore the space of states which could reach the goal by performing any transitions. Thus, starting from the target one explores possible paths to the target in a *backwards* manner.

At its core lies the *MaxU* operation for computing maximal probabilities to fulfil a PTCTL until-formula. As we consider the special case of reachability problems, we will not define this operation for general until-formulae but only for the special case of reachability targets. The algorithm transforms symbolic states to obtain symbolic states that represent all potential predecessors of a given delay or discrete transition. To bridge between targets, which are just collections of states, and symbolic states, we define symbolic target assignments, which group valuations by location.

$$\frac{}{\swarrow(T) \subseteq S_{\mathsf{bwd}}} \text{ (BASE)} \qquad \frac{\begin{array}{c}(l,\zeta) \in S_{\mathsf{bwd}} \quad \bar{e} = (l',\alpha, R, l) \\ \mathsf{trans}[l',\alpha](R,l) \neq 0 \\ \emptyset \neq (l',\zeta') = \mathsf{pre}_{\bar{e}}(\mathsf{pre}_{time}(l,\zeta)) \not\sqsubseteq \swarrow(T)\end{array}}{(l',\zeta') \in S_{\mathsf{bwd}} \quad (l',\zeta') \underset{\langle R,l\rangle}{\overset{(l',\alpha)}{\text{-bwd}\rightarrow}} (l,\zeta)} \text{ (STEP)}$$

$$\frac{(l,\zeta_i) \underset{\langle R_i,l_i\rangle}{\overset{e}{\text{-bwd}\rightarrow}} - \quad (R_1,l_1) \neq (R_2,l_2) \quad \zeta_1 \cap \zeta_2 \neq \emptyset \quad (l,\zeta_1 \cap \zeta_2) \not\sqsubseteq \swarrow(T)}{(l,\zeta_1 \cap \zeta_2) \in S_{\mathsf{bwd}}} \text{ (SUB)}$$

Figure 4: The rules used to build the symbolic backwards system, based on the MaxU-algorithm by [64].

**Definition 6.4** (Target assignment)**.** Given a pPpTA $\mathcal{P}$ and target $T \subseteq \mathrm{Loc} \times \mathrm{Val}(\mathrm{Clk})$. One can interpret $T$ equivalently as a *target assignment* $T\colon \mathrm{Loc} \to \mathcal{P}(\mathrm{Val}(\mathrm{Clk})), l \mapsto \{\tau \in \mathrm{Val}(\mathrm{Clk}) \mid (l,\tau) \in F\}$.

$T$ is a *symbolic target assignment* if $(l, T(l) \times \mathrm{Val}(\mathrm{P_{Clk}}))$ is a symbolic state for $l \in \mathrm{Loc}$, and $T$ is a *target zone assignment*, if $(l, T(l) \times \mathrm{Val}(\mathrm{P_{Clk}}))$ is a zone for $l \in \mathrm{Loc}$.

That the backwards exploration approach can be lifted to probability-parametric probabilistic timed automata was shown in [40]. Further, extending this to the clock-parametric setting is not difficult, just technically involved. As there are slight differences in the algorithm between the conference and journal version of [64] and recent replication studies [41] highlighted hidden difficulties for implementation, we will present the adapted algorithm with some more details.

**Definition 6.5** (Symbolic backwards system)**.** Given a pPpTA $\mathcal{P}$. For symbolic target assignment $T$, let

$$\swarrow(T) \coloneqq \{\mathsf{pre}_{time}(l, T(l) \times \mathrm{Val}(\mathrm{P_{Clk}})) \mid l \in \mathrm{Loc} \text{ with } T(l) \neq \emptyset\}.$$

We define the *symbolic backwards system* $\mathsf{Sys}_{\mathcal{P}}^{\mathsf{bwd}}(T)$ as the smallest labelled transition system $(S_{\mathsf{bwd}}, \text{-bwd}\rightarrow)$ closed under the rules (BASE), (STEP) and (SUB) given in Figure 4. Here, $(l,\zeta) \sqsubseteq \swarrow(T)$ denotes that $\zeta \subseteq \zeta'$ for some $(l,\zeta') \in \swarrow(T)$.

Rule (BASE) ensures that all target states are in the system. The (STEP) rule performs combined backwards steps for every discovered symbolic state. This is done by first finding all time delay predecessors of the symbolic state. Every possible concrete action is examined to check whether it could be performed in a forward execution. This is achieved by verifying for $l' \in \mathrm{Loc}$, $\alpha \in \mathrm{Act}$ and $R \subseteq \mathrm{Clk}$ whether the transition kernel could forward non-zero (possibly parametric) probability mass to the current location $l$ (condition $\mathsf{trans}[l',\alpha](R,l) \neq 0$). Such possible transitions are explored backwards from the time predecessors of the current symbolic states. The so found states, if there are any, make up another symbolic state in the backwards system. There is an additional check $(l', \zeta') \not\sqsubseteq \swarrow(T)$, to prevent further exploration if one discovers states which can reach the target simply by performing time delays. Since for such states an optimal scheduler can just select the appropriate time delay to certainly reach the target.

The third rule (SUB) is used to combine symbolic states which reached the same location via different outcomes, i.e. different clock resets and/or target location, of the same action.

If we let $\Phi_T$ denote the monotone operator which, given a labelled transition system $(S, \rightarrow)$, extends it to the labelled transition system $(S', \rightarrow')$ with $S \subseteq S', \rightarrow \subseteq \rightarrow'$ which adds all states/transitions that can be derived in a single step using one of the above rules from $(S, \rightarrow)$, we obtain

$$\mathsf{Sys}^{\mathsf{bwd}}_{\mathcal{P}}(T) = \bigcup_{k \in \mathbb{N}_0} \Phi^k_T(\emptyset, \emptyset).$$

In particular, every application of $\Phi_T$ can only add a finite number of new states and transitions, thus $S_{\mathsf{bwd}}$ and $\text{-bwd} \rightarrow$ are countable.

The transitions in the backwards system represent combined steps backwards in both time and through a discrete labelled action. Throughout the exploration multiple such edges coming from different probabilistic choices of the same labelled transition will be discovered. These need to be recombined with an additional operation to obtain a pMDP.

**Definition 6.6** (Maximal edge selection). Given a pPpTA $\mathcal{P}$, symbolic target assignment $T$ and the symbolic backwards system $\mathsf{Sys}^{\mathsf{bwd}}_{\mathcal{P}}(T) = (S_{\mathsf{bwd}}, \text{-bwd} \rightarrow)$. A subsystem $M = (S', \longrightarrow')$ is a labelled transition system with $S' \subseteq S_{\mathsf{bwd}}$ and $\longrightarrow' \subseteq \text{-bwd} \rightarrow$.

In $M$ a set of edges $\mathcal{E} \subseteq \longrightarrow', \mathcal{E} \neq \emptyset$ is called a *maximal selection of edges for $\alpha \in \text{Act}$ at* $(l, \zeta) \in S'$ if it fulfils:

i) $\mathcal{E} \subseteq (l, -) \xrightarrow{\alpha}' - = \{ (l, \zeta_1) \xrightarrow[\langle R, l' \rangle]{\alpha}{}' (l', \zeta_2) \mid l' \in \text{Loc}, R \subseteq \text{Clk}, \zeta_1, \zeta_2 \subseteq \text{Val}(\text{Clk}, \text{P}_{\text{Clk}}) \}$,

ii) for $(l, \zeta') \xrightarrow{\alpha}' - \in \mathcal{E}: \zeta \subseteq \zeta'$,

iii) for $e_i = (l, \zeta_i) \xrightarrow[\langle R_i, l_i \rangle]{\alpha}{}' - \in \mathcal{E}$ with $e_1 \neq e_2: (R_1, l_1) \neq (R_2, l_2)$, and

iv) $\mathcal{E}$ is maximal w.r.t. $\subseteq$.

We let $MaxSel_M(\alpha, (l, \zeta))$ denote the set of all maximal edge selections for $\alpha$ at $(l, \zeta)$ in $M$. For $\mathcal{E} \in MaxSel_M(\alpha, (l, \zeta))$ we define the partial function $step_{\mathcal{E}} : \mathcal{P}(\text{Clk}) \times \text{Loc} \rightharpoonup S'$ with $step_{\mathcal{E}}(R', l') = (l', \zeta')$ for the unique $(l, -) \xrightarrow[\langle R', l' \rangle]{\alpha}{}' (l', \zeta') \in \mathcal{E}$ if it exists.

A maximal edge selection for $\alpha$ at $(l, \zeta)$ contains at most one concrete representative symbolic state $(l', \zeta')$ for every possible choice $(R, l')$ of clocks $R$ to reset and location $l'$ to transition to, which represents possible states that could be reached when performing discrete action $\alpha$ from states in $(l, \zeta)$.

The maximality condition for the maximal edge selections ensures that the probabilistic transitions assembled from the selections allow to maximize the probability of reaching the target. Transitions build from maximal edge selections would always be preferred by a maximizing scheduler over transitions build from non-maximal edge selections. Hence, we leave out non-maximal edge selections from the construction, as they bring no further benefit.

Note that $step_{\mathcal{E}}$ is measurable since its domain is finite and in the associated power-set $\sigma$-algebra every set is measurable. We can now build a pMDP over the symbolic states found in the backwards system by building (parametric) probabilistic transitions using the maximal edge selections.

**Definition 6.7.** Given a p$\mathbb{P}$pTA $\mathcal{P}$, its concrete semantics symbolic target assignment $T$ and a subsystem $M = (S', \longrightarrow')$ of the symbolic backward system $\mathsf{Sys}_{\mathcal{P}}^{\mathsf{bwd}}(T)$. The associated sub-pMDP $\langle M \rangle := (S_{\langle M \rangle}, L_{\langle M \rangle}, \mathrm{trans}_{\langle M \rangle})$ consists of

- states $S_{\langle M \rangle} = S'$ from $M$,
- labels $L_{\langle M \rangle} = \{ (\alpha, \mathcal{E}) \mid \alpha \in \mathrm{Act}, (l, \zeta) \in S_{\langle M \rangle}, \mathcal{E} \in MaxSel_M(\alpha, (l, \zeta)) \}$, and
- transitions $\mathrm{trans}_{\langle M \rangle}[(l, \zeta), (\alpha, \mathcal{E})] := step_{\mathcal{E}} \# \mathrm{trans}[(l, \alpha)]$ for $(l, \zeta) \in S_{\langle M \rangle}$, $\alpha \in \mathrm{Act}$ and $\mathcal{E} \in MaxSel_M(\alpha, (l, \zeta))$.

Note that $S_{\langle M \rangle} = S'$ is countable, as established before. Importantly, $L_{\langle M \rangle}$ is also countable: The set of maximal edge selections $MaxSel_M(\alpha, (l, \zeta))$ is countable for every $\alpha \in \mathrm{Act}, (l, \zeta) \in S_{\langle M \rangle}$, because every action label represents a (parametric) probabilistic transition whose domain are pairs of clock resets and locations. As both of these are finite, every maximal edge selection consists of finitely many symbolic backwards edges, selected from a (subset of a) countable set $\longrightarrow'$. As finite subsets of countable sets are countable, $MaxSel_M(\alpha, (l, \zeta))$ is countable. Because there are only finitely many action labels in Act and countable many symbolic states in $S_{\langle M \rangle}$, $L_{\langle M \rangle}$ is countable. Hence, we use the power set as the associated $\sigma$-algebra for both $S_{\langle M \rangle}$ and $L_{\langle M \rangle}$.

The so constructed sub-pMDP is similar to the p$\mathbb{P}$pTA's actual semantics, but it hides the progress of time into the symbolic states. This is sensible, since time delay transitions are always deterministic and lead to a certain next state with probability one. Thus, if there is some *optimal* labelled action reachable via delays, an optimizing scheduler can always choose to wait the desired time until this action becomes enabled to ensure optimal reachability probabilities. The uncertain decisions which need to be optimized by a scheduler thus mainly concern the labelled actions of the p$\mathbb{P}$pTA.

**Definition 6.8** (Backwards semantics)**.** Given a p$\mathbb{P}$pTA $\mathcal{P}$ and symbolic target assignment $T$. We define the *symbolic backwards semantics* $[\![\mathcal{P}]\!]_{\mathsf{bwd}}^T$ as the sub-pMDP $\langle \mathsf{Sys}_{\mathcal{P}}^{\mathsf{bwd}}(T) \rangle$.

Note that all operations involved in the definition of the backwards semantics preserve zones, i.e. if one starts the construction with target zone assignment $T$, all constructed symbolic states will be zones. This allows for an efficient implementation of the approach using parametric Difference Bounded Matrices [45] as a representation of zones.

Since p$\mathbb{P}$pTA subsume pTA, for which the reachability problem is already undecidable, the construction of $[\![\mathcal{P}]\!]_{\mathsf{bwd}}^T$ will not terminate in many cases. One solution to this is limiting the clock parameter valuations to a finite set. In this case this backwards semantics is a combination of finitely many backwards semantics on p$\mathbb{P}$TA of which each is finite by [64].

Alternatively, we can also use the backwards exploration on the whole parameter valuation space to obtain bounds on the reachability probabilities: Stopping the backwards exploration after finitely many steps produces a subsystem which can be used to lower bound the correct reachability probability.

**Proposition 6.9.** *Given a pPpTA $\mathcal{P}$, its concrete semantics $[\![\mathcal{P}]\!] = (\mathbb{S}, \mathrm{Act} \,\dot\cup\, \mathbb{R}_{\geq 0}, step)$ and a symbolic target assignment $T$.*

  i) *For any subsystem $M$ of the symbolic backward system $\mathsf{Sys}_\mathcal{P}^{\mathsf{bwd}}(T)$, symbolic state $(l, \zeta) \in S_{(\!|M|\!)}$ and $(l, \nu) \in \mathsf{pre}_{time}(l, \zeta)$:*

$$\overline{\mathbb{Pr}}_{(l,\zeta)}^{(\!|M|\!)}[\rho](\Diamond{\swarrow}(T)) \leq \overline{\mathbb{Pr}}_{\mathfrak{d}_{(l,\nu)}}^{[\![\mathcal{P}]\!]}[\rho](\blacklozenge T)$$

  *for $\rho \in \mathrm{Val}_\mathcal{P}(\mathrm{P}_{prob})$.*

  ii) *For any concrete state $(l, \nu) \in \mathbb{S}$ with $\overline{\mathbb{Pr}}_{\mathfrak{d}_{(l,\nu)}}^{[\![\mathcal{P}]\!]}[\cdot](\blacklozenge T) \neq 0$ there exists a symbolic state $(l, \zeta) \in S_{(\!|\mathsf{Sys}_\mathcal{P}^{\mathsf{bwd}}(T)|\!)}$ with $(l, \nu) \in \mathsf{pre}_{time}(l, \zeta)$, and*

$$\overline{\mathbb{Pr}}_{\mathfrak{d}_{(l,\nu)}}^{[\![\mathcal{P}]\!]}[\rho](\blacklozenge T) \leq \overline{\mathbb{Pr}}_{(l,\zeta)}^{(\!|\mathsf{Sys}_\mathcal{P}^{\mathsf{bwd}}(T)|\!)}[\rho](\Diamond{\swarrow}(T))$$

  *for $\rho \in \mathrm{Val}_\mathcal{P}(\mathrm{P}_{prob})$.*

The proofs of these statements, like the whole construction, follow [64]. One reasons by induction about reachability within a limited number of steps, i.e. that the target must be reached after at most $k \in \mathbb{N}_0$ discrete transitions ($\blacklozenge^{\leq k} T$). These are precisely the states discovered by the backwards exploration operations $\Phi_T^k$. Corresponding choices for the actions that the schedulers need to pick to reach the target are also immediately constructed during the exploration. For the other direction, every concrete state that can reach the target with non-zero probability, must do so for the first time after finitely many steps. Hence, a corresponding backwards exploration will also find such a path after finitely many iterations.

Formalizing this reasoning involves the underlying parametric Markov Chains created by schedulers on the corresponding pMDPs, restricted to valid probability parameter valuations. The only difference to the original proof lies in the technical necessity to consider only valid valuations for the probability parameters so that we actually obtain Markov Chains in every step.

**Theorem 6.10.** *Given a pPpTA $\mathcal{P}$, a state $(l, \nu) \in \mathbb{S}$ in its concrete semantics $[\![\mathcal{P}]\!] = (\mathbb{S}, \mathrm{Act} \,\dot\cup\, \mathbb{R}_{\geq 0}, step)$, and symbolic target assignment $T$.*

  i) *There exists a symbolic state $(l, \zeta) \in S_{(\!|\mathsf{Sys}_\mathcal{P}^{\mathsf{bwd}}(T)|\!)}$ with $(l, \nu) \in \mathsf{pre}_{time}(l, \zeta)$ if and only if $\overline{\mathbb{Pr}}_{\mathfrak{d}_{(l,\nu)}}^{[\![\mathcal{P}]\!]}[\cdot](\blacklozenge T) \neq 0$.*

  ii) *If $\overline{\mathbb{Pr}}_{\mathfrak{d}_{(l,\nu)}}^{[\![\mathcal{P}]\!]}[\cdot](\blacklozenge T) \neq 0$ then:*

$$\overline{\mathbb{Pr}}_{\mathfrak{d}_{(l,\nu)}}^{[\![\mathcal{P}]\!]}[\cdot](\blacklozenge T) = \max_{\substack{(l,\zeta) \in S_{(\!|\mathsf{Sys}_\mathcal{P}^{\mathsf{bwd}}(T)|\!)} \\ (l,\nu) \in \mathsf{pre}_{time}(l,\zeta)}} \overline{\mathbb{Pr}}_{(l,\zeta)}^{(\!|\mathsf{Sys}_\mathcal{P}^{\mathsf{bwd}}(T)|\!)}[\cdot](\Diamond{\swarrow}(T)).$$

Note that in the first part the existence of a suitable symbolic state $(l, \zeta) \in S_{(\!|\mathsf{Sys}_\mathcal{P}^{\mathsf{bwd}}(T)|\!)}$ only gives a path via symbolic states to the symbolic target ${\swarrow}(T)$. Special care needs to

be taken when constructing a corresponding actual path for the concrete state $(l, \nu)$ to $T$ to ensure all parameters can be properly instantiated so that the target is reachable with non-zero probability. But even if there are some instantiations that do not lead to proper paths, these play no further role in the second part of the theorem which only concerns valid parameter valuations.

The notation suggests that the given correctness statement is just a point-wise lifting of the corresponding statement on non-probability parametric $\mathbb{P}$pTAs for every valid probability parameter valuation $\rho \in \mathrm{Val}_\mathcal{P}(\mathrm{P}_{prob})$. In particular, the point-wise lifting would correspond to the statement, that for every probability parameter valuation we can pick a suitable symbolic state which gives the same maximal reachability probability. The actual statement is stronger in that *there is a single symbolic state that attains the maximal reachability probability for every possible probability parameter valuation*.

However, it turns out both statements are equivalent, due to the commutativity of the backwards reachability semantics with probability parameter instantiation: every state in $\mathsf{Sys}^{\mathsf{bwd}}_{\mathcal{P}[\rho]}(T)$ is identically present in $\mathsf{Sys}^{\mathsf{bwd}}_{\mathcal{P}}(T)$, but the latter might contain even more states. This is because first instantiating probability parameters might cause $\mathrm{trans}_\rho[l, \alpha]$ to be zero, thus eliminating the transition from exploration, while its parametric version $\mathrm{trans}[l, \alpha]$ needs to be further explored. Consequently, certain arcs will not be explored backwards, thus potentially reducing the state space of $\mathsf{Sys}^{\mathsf{bwd}}_{\mathcal{P}[\rho]}(T)$. But the additionally discovered states in $\mathsf{Sys}^{\mathsf{bwd}}_{\mathcal{P}}(T)$ play no role, since the parametric arcs that led to the states' discovery will be set to zero in $[\![\mathcal{P}]\!]^T_{\mathsf{bwd}}[\rho]$. Hence, no difference in reachability probabilities for target $T$ occurs between $[\![\mathcal{P}[\rho]]\!]^T_{\mathsf{bwd}}$ and $[\![\mathcal{P}]\!]^T_{\mathsf{bwd}}[\rho]$.

The observation that for every concrete state $(l, \nu)$ there is a *globally maximizing* symbolic state, independent of the choice of the probabilistic parameters, can be used to separate the optimal parameter synthesis for maximal reachability objectives into two steps. We can consider all symbolic states which contain a concrete initial state $(l_0, \underline{0} \,\dot\cup\, \gamma)$. From these one picks the symbolic state $(l_0, \zeta_{opt})$ with the maximal probability of reaching the target. The clock-parameter valuations of all initial states contained in this optimal symbolic state form a set of equivalent optimal solutions.

**Corollary 6.11** (Preservation of maximal reachability probabilities)**.** *Given a $p\mathbb{P}pTA$ $\mathcal{P}$ and symbolic target assignment $T$. For $\Theta_{\mathrm{Clk}} \subseteq \mathrm{Val}_\mathcal{P}(\mathrm{P}_{\mathrm{Clk}})$, $\Theta_{prob} \subseteq \mathrm{Val}_\mathcal{P}(\mathrm{P}_{prob})$:*

$$\overline{\mathbb{Pr}}^{\mathcal{P}}_{\max \Theta_{\mathrm{Clk}}, \Theta_{prob}}(\blacklozenge T) = \max_{(l_0, \zeta) \in \mathcal{S}^{\Theta_{\mathrm{Clk}}}_0} \overline{\mathbb{Pr}}^{(\!|\mathsf{Sys}^{\mathsf{bwd}}_{\mathcal{P}}(T)|\!)}_{(l_0, \zeta), \max \Theta_{prob}}(\lozenge \swarrow(T)),$$

*with*

$$\mathcal{S}^{\Theta_{\mathrm{Clk}}}_0 := \{\, (l_0, \zeta) \in S_{(\!|\mathsf{Sys}^{\mathsf{bwd}}_{\mathcal{P}}(T)|\!)} \mid \exists \gamma \in \Theta_{\mathrm{Clk}} \colon (l_0, \underline{0} \,\dot\cup\, \gamma) \in \mathsf{pre}_{time}(l_0, \zeta) \,\}.$$

In a second step one can then determine optimizing probability parameter valuations. Starting from the determined optimizing symbolic state $(l_0, \zeta^{opt}_0)$ in $[\![\mathcal{P}[\rho]]\!]^T_{\mathsf{bwd}}$, which is a countable pMDP, one can apply common methods developed for pMDP's to synthesize optimizing parameter values for the probability parameters.

This is also the core idea of the clock parameter equivalence class partitioning of [13] to obtain a valuation for clock parameters that gives maximal reachability probabilities. By repeating this construction with $\mathcal{S}^{\Theta_{\mathrm{Clk}}}_{i+1} := \{\, (l_0, \zeta \setminus \zeta^{opt}_0) \mid (l_0, \zeta) \in \mathcal{S}^{\Theta_{\mathrm{Clk}}}_i \,\} \setminus \{\emptyset\}$ one can get the $i$-th-best set of clock parameter valuations. Iterating this procedure thus gives a suitable partitioning of the clock parameter state space.

## 6.3 Digital Clocks

Another way to abstract the uncountable state space introduced by dense-time clocks is through discretization of the time domain. The most common way is to restrict to integer time-steps, resulting in the digital clock semantics [43]. The digital clock semantics was considered for probabilistic timed automata by [61] and extended to probability-parameters in [40]. As before with the backwards semantics, we will see that an extension to clock parameters is also possible in a straight-forward manner. The same restriction as in the non-parametric cases applies, namely that this method will only be correct for parametric timed automata which do not contain strict inequalities in constraints.

**Definition 6.12** (Closed pℙpTA). A pℙpTA $\mathcal{P}$ is called *non-strict* or *closed*, if no $\varphi \in$ Im (inv) ∪ Im (guard) contains $c_1 \bowtie t$ or $c_1 - c_2 \bowtie t$ for any $c_1, c_2 \in \mathrm{Clk}, t \in \mathbb{N}_0 \,\dot{\cup}\, \mathrm{P}_{\mathrm{Clk}}, \bowtie \in \{<, >\}$, i.e. there are no strict inequalities in any constraint.

Digital clock semantics do not require the definition of a new model to represent the abstracted semantics. Instead, the behaviour is readily available in the concrete semantics $[\![\mathcal{P}]\!]$ through selection of appropriate schedulers which only pick integer delays for time delay transitions.

**Definition 6.13** (Integral-time-step scheduler). Given a pℙpTA $\mathcal{P}$. An *integral-time-step scheduler* $\sigma\colon \mathsf{Paths}^{<\omega}_{[\![\mathcal{P}]\!]} \rightsquigarrow_{\mathbb{Q}(\mathrm{P}_{prob})} \mathrm{Act} \,\dot{\cup}\, \mathbb{R}_{\geq 0}$ is a scheduler with $\sigma[\pi](\mathbb{R}_{\geq 0} \setminus \mathbb{N}_0) = 0$, or equivalently $|\sigma_\rho[\pi]| = \sigma_\rho[\pi](\mathrm{Act} \,\dot{\cup}\, \mathbb{N}_0)$ for $\rho \in \mathrm{Val}_{\mathcal{P}}(\mathrm{P}_{prob})$. The set of all integral-time-step schedulers of $[\![\mathcal{P}]\!]$ is denoted by $\mathrm{Sched}^{\mathbb{N}_0}_{[\![\mathcal{P}]\!]}$.

Stated differently, an integral-time-step scheduler almost-surely selects labelled actions or integer time delays, i.e. no non-integer delays. As the initial state always has all clocks set to zero, starting from an integer clock valuation ensures that all states reached under such a scheduler only contain integer clock values. This allows to effectively reduce the state space from a dense uncountable base set ($\mathbb{R}_{\geq 0}$) to a countable one ($\mathbb{N}_0$).

With the change of the time domain for digital clock semantics, we could wrongly claim to have visited a target with non-integer clock values due to the definition of $\blacklozenge F$. Hence, we restrict our attention to reachability objectives that only try to reach a certain location and do not concern clock valuations.

**Definition 6.14.** Given a pℙpTA $\mathcal{P}$. The *target states at locations* $L \subseteq \mathrm{Loc}$ *with times in* $\mathbb{T} \in \{\mathbb{N}_0, \mathbb{R}_{\geq 0}\}$ is defined as the target $L@\mathbb{T} := \{(l, \tau) \in L \times \mathrm{Val}(\mathrm{Clk}) \mid \mathrm{Im}(\tau) \subseteq \mathbb{T}\}$.

The intuitive reason why integral-time-step schedulers are sufficient to obtain optimal reachability probabilities in closed pℙpTAs lies in the core idea that, without strict inequalities in constraints, it can never occur that getting infinitesimally close to a constraint's bound results in better reachability probabilities, than just reaching the bound and continuing from there. This is because potential infinitesimal savings made in earlier delay transitions cannot be used to fulfil a constraint at a later point during the execution that would otherwise become unsatisfied.

To prove that integral-time-step schedulers are sufficient, we want to convert a normal scheduler into an integral-time-step one while retaining the same reachability probabilities, up to changes of clock values induced by the conversion. This is done by rounding the clock valuations along a path induced by the scheduler to integers, depending on a threshold $\varepsilon$.

**Definition 6.15** ($\varepsilon$-digitization). For $t \in \mathbb{R}, \varepsilon \in [0,1]_\mathbb{R}$ let:

$$[t]_\varepsilon = \begin{cases} \lfloor t \rfloor, & \text{if } t \leq \lfloor t \rfloor + \varepsilon, \\ \lceil t \rceil, & \text{else.} \end{cases}$$

Given a closed pPpTA $\mathcal{P}$, the $\varepsilon$-*digitization of path* $\pi \in \mathsf{Paths}_{\llbracket \mathcal{P} \rrbracket}$ is the path $[\pi]_\varepsilon \in \mathsf{Paths}_{\llbracket \mathcal{P} \rrbracket}$ with $|\pi| = |[\pi]_\varepsilon|$ and

- $[\pi]_\varepsilon[i] = ([\tau]_\varepsilon, \gamma)$, where $\pi[i] = (\tau, \gamma)$ and for $c \in \mathrm{Clk}$: $[\tau]_\varepsilon(c) = [\mathrm{dur}(\pi, i)]_\varepsilon - [\mathrm{dur}(\pi, j_c)]_\varepsilon$ with $j_c \in [0, i]$ such that $\tau(c) = \mathrm{dur}(\pi, i) - \mathrm{dur}(\pi, j_c)$.

- $\mathrm{act}_i([\pi]_\varepsilon) = \begin{cases} \mathrm{act}_i(\pi), & \text{if } \mathrm{act}_i(\pi) \in \mathrm{Act}, \\ [\mathrm{dur}(\pi, i)]_\varepsilon - [\mathrm{dur}(\pi, i-1)]_\varepsilon, & \text{if } \mathrm{act}_i(\pi) \in \mathbb{R}_{\geq 0}. \end{cases}$

The definition is the straight-forward generalization of the definitions by [43, 61] with clock parameter valuations remaining unaffected by the rounding, as they already only map to natural numbers anyway. Note that, e.g., $[\cdot]_\varepsilon \colon \mathsf{Paths}^\omega_{\llbracket \mathcal{P} \rrbracket} \to \mathsf{Paths}^\omega_{\llbracket \mathcal{P} \rrbracket}$ is a measurable function with $\mathrm{Im}([\cdot]_\varepsilon) \subseteq (\mathbb{S} \times (\mathrm{Act} \mathbin{\dot\cup} \mathbb{N}_0))^\omega$. Because there is a certain interval, depending only on $\varepsilon$, around each integer in which every number is mapped to this integer by the $\varepsilon$-digitization, and intervals are the basic measurable sets in the Borel-$\sigma$-algebra of $\mathbb{R}_{\geq 0}$.

**Lemma 6.16.** *Given a closed pPpTA $\mathcal{P}$. For every scheduler $\sigma \colon \mathsf{Paths}^{<\omega}_{\llbracket \mathcal{P} \rrbracket} \rightsquigarrow_{\mathbb{Q}(\mathrm{P}_{prob})} \mathrm{Act} \mathbin{\dot\cup} \mathbb{R}_{\geq 0}$ and $\varepsilon \in [0,1]_\mathbb{R}$, there exists an integral-time-step scheduler $\sigma_\varepsilon \colon \mathsf{Paths}^{<\omega}_{\llbracket \mathcal{P} \rrbracket} \rightsquigarrow_{\mathbb{Q}(\mathrm{P}_{prob})} \mathrm{Act} \mathbin{\dot\cup} \mathbb{R}_{\geq 0}$ such that*

$$\mathbb{Pr}^{\mathcal{P}}_{\gamma, \sigma_\varepsilon} = [\cdot]_\varepsilon \# \mathbb{Pr}^{\mathcal{P}}_{\gamma, \sigma}.$$

The construction is technically involved but analogous to the non-parametric version given in [61]. The introduction of clock parameters does not influence the possible $\epsilon$-digitization of paths/schedulers, since the parameter values remain unchanged on every possible path. Also the introduction of probability parameters also has no direct effect on the construction, as the paths themselves remain unaffected. Solely the type of the objects changes, now being parametric distributions/kernels instead of normal ones.

**Theorem 6.17.** *Let $\mathcal{P}$ be a closed pPpTA and $L \subseteq \mathrm{Loc}$ be a set of target locations. For all $\gamma \in \mathrm{Val}_\mathcal{P}(\mathrm{P}_{\mathrm{Clk}})$ and $\rho \in \mathrm{Val}_\mathcal{P}(\mathrm{P}_{prob})$:*

$$\overline{\mathbb{Pr}}^{\mathcal{P}}_\gamma[\rho](\blacklozenge L @ \mathbb{R}_{\geq 0}) = \sup_{\sigma \in \mathsf{Sched}_{\llbracket \mathcal{P} \rrbracket}} \mathbb{Pr}^{\llbracket \mathcal{P} \rrbracket}_{\gamma, \sigma}[\rho](\blacklozenge L @ \mathbb{R}_{\geq 0}) = \sup_{\sigma \in \mathsf{Sched}^{\mathbb{N}_0}_{\llbracket \mathcal{P} \rrbracket}} \mathbb{Pr}^{\llbracket \mathcal{P} \rrbracket}_{\gamma, \sigma}[\rho](\blacklozenge L @ \mathbb{N}_0),$$

*and*

$$\underline{\mathbb{Pr}}^{\mathcal{P}}_\gamma[\rho](\blacklozenge L @ \mathbb{R}_{\geq 0}) = \inf_{\sigma \in \mathsf{Sched}_{\llbracket \mathcal{P} \rrbracket}} \mathbb{Pr}^{\llbracket \mathcal{P} \rrbracket}_{\gamma, \sigma}[\rho](\blacklozenge L @ \mathbb{R}_{\geq 0}) = \inf_{\sigma \in \mathsf{Sched}^{\mathbb{N}_0}_{\llbracket \mathcal{P} \rrbracket}} \mathbb{Pr}^{\llbracket \mathcal{P} \rrbracket}_{\gamma, \sigma}[\rho](\blacklozenge L @ \mathbb{N}_0).$$

For reachability of target locations, we get that it suffices to consider only integral-time-step schedulers, reducing the relevant state space to a countable size. To further shrink the space to a finite domain, one can consider an upper-bound on clock parameter values [17, 48]. This allows to establish a maximal integer against which a clock's value could ever be compared during an execution. Any clock values above this threshold do not change the satisfaction of the constraints, similar to how decreasing/increasing parameter values in L/U-automata has no observable effect beyond a certain threshold. Hence, such valuations can be combined into an abstract state without influencing the semantics of the modelled pPpTA. Together with the integer valued clocks, this then allows for a finite state space which could be explored exhaustively.

# 7 Conclusion

We presented probability-parametric and clock-parametric probabilistic timed automata (pPpTA) as a formalism to combine parametric modelling for both quantitative aspects of probabilistic timed automata. The model is a straight-forward generalization of the models introduced in [40, 47] which only supported one kind of parameters. By separating the set of parameters for probabilities and clocks we can handle each type of parameter independently. We focused on the problem of synthesizing optimal parameter valuations for maximal/minimal reachability objectives. In our treatment we focused on clock parameters and showed how many existing approaches from the literature on non-parametric probabilistic timed automata can be lifted to this more general setting.

We showed that the concept of L/U-automata by [45] can be extended with probability parameters. As one result, one can directly determine optimal valuations and eliminate certain clock parameters from in the case of L/U-automata, and thus reduce the problem to the known case of probability-parametric $\mathbb{P}$TAs.

We presented generalizations of the backwards semantics [64] and digital clock semantics [43] to abstract the state space of the pPpTA-model. There exists a wide variety of other approaches for the analysis of reachability problems on (probabilistic) timed automata, e.g. a forwards semantics [30, 73] analogous to the backwards semantics taking combined discrete and time delay transitions, but starting from the initial states going forwards. It was already established in [62] that such an approach does not give exact reachability probabilities for probabilistic timed automata. Instead, this semantics can only be used to obtain upper bounds, which unfortunately can be arbitrarily loose. We believe that an analogous extension to pPpTAs is possible, but we did not present it here as it would not provide interesting insights.

Newer approaches to a forwards analysis of probabilistic timed automata involve stochastic games and have been presented for $\mathbb{P}$TAs in [60] and for $\mathbb{P}$pTAs in [47]. Lifting this approach to probability-parametric models seems like a reasonable endeavour, the resulting abstract model being a parametric stochastic game. While theoretically defining this model is certainly possible, right now, a huge body of research as for parametric Markov Decision Processes is still missing. Consequently, such an abstraction can currently not benefit from many results from the literature. Thus, investigating this approach and other methods, e.g. inverse methods [8, 7], is left open for future work.

A related generalization of existing formalisms is given in [71] with clock-dependent $\mathbb{P}$TAs. Rather than introducing parameters for probabilities or clocks, the transition probabilities in this model can depend on the current clock valuation. While this enables to model more intricate behaviour, it also significantly complicates analysis of the model. As there is no strict separation between the timed and probabilistic behaviour, a sequential analysis of the two aspects, as done in this work, is not applicable in an obvious way. Considering how and whether the approaches presented in this work can be applied to this (or an even more) general model is also interesting for future work.